\documentclass[aps,prl,twocolumn,10pt,longbibliography]{revtex4-1}
\usepackage{amsmath}
\usepackage{amsfonts}
\usepackage{amssymb}
\usepackage{bm}
\usepackage{epsfig}
\usepackage{color}
\usepackage{graphicx}
\usepackage[hyperindex=true]{hyperref}
\hypersetup{linktocpage,colorlinks=true,citecolor=blue,linkcolor=blue}
\usepackage[inline]{enumitem}
\usepackage{braket}
\usepackage[normalem]{ulem}
\usepackage{pdfpages}
\usepackage{pgffor}

\makeatletter
\AtBeginDocument{\let\LS@rot\@undefined}

\makeatletter
\def\hlinewd#1{%
\noalign{\ifnum0=`}\fi\hrule \@height #1 %
\futurelet\reserved@a\@xhline}
\makeatother

\makeatother

\begin{document}

\title{Fluctuating Forces Induced by Non Equilibrium and Coherent Light Flow}

\author{Ariane Soret$^{1,2}$, Karyn Le Hur$^2$, Eric Akkermans$^{1}$}
\email{eric@physics.technion.ac.il}
\affiliation{$^1$ Department of Physics, Technion -- Israel Institute of Technology, Haifa 3200003, Israel}
\affiliation{$^2$Centre de Physique Th\'eorique, \'Ecole Polytechnique, CNRS, Universit\'e Paris-Saclay, 91128 Palaiseau, France}

\begin{abstract}
We show that mesoscopic coherent fluctuations of light propagating in random media induce fluctuating radiation forces. A hydrodynamic Langevin approach is used to describe the coherent light fluctuations, whose noise term  accounts for mesoscopic coherent effects. This description -- generalizable to other quantum or classical wave problems -- allows to understand coherent fluctuations as a non equilibrium light flow, characterized by the diffusion coefficient $D$ and the mobility $\sigma$, otherwise related by a Einstein relation. The strength of these fluctuating forces is determined by a single dimensionless and tunable parameter, the conductance $g_{\mathcal{L}}$. Orders of magnitude of these fluctuation forces are offered which show experimental feasibility.
\end{abstract}

\maketitle

Casimir physics covers a wealth of phenomena where forces between macroscopic objects are induced by long range fluctuations \cite{Kardar99} of either classical or quantum origin.  Fluctuations of the quantum electrodynamic (QED) vacuum epitomize this type of physics \cite{Casimir48}, but such fluctuation induced forces (FIF) arise in a wide range of systems \cite{DeGennes78,Kafri15,Kirkpatrick14,Dean16,Reynaud09}. 

\begin{figure}[t]
\centering
\includegraphics[width=1\columnwidth]{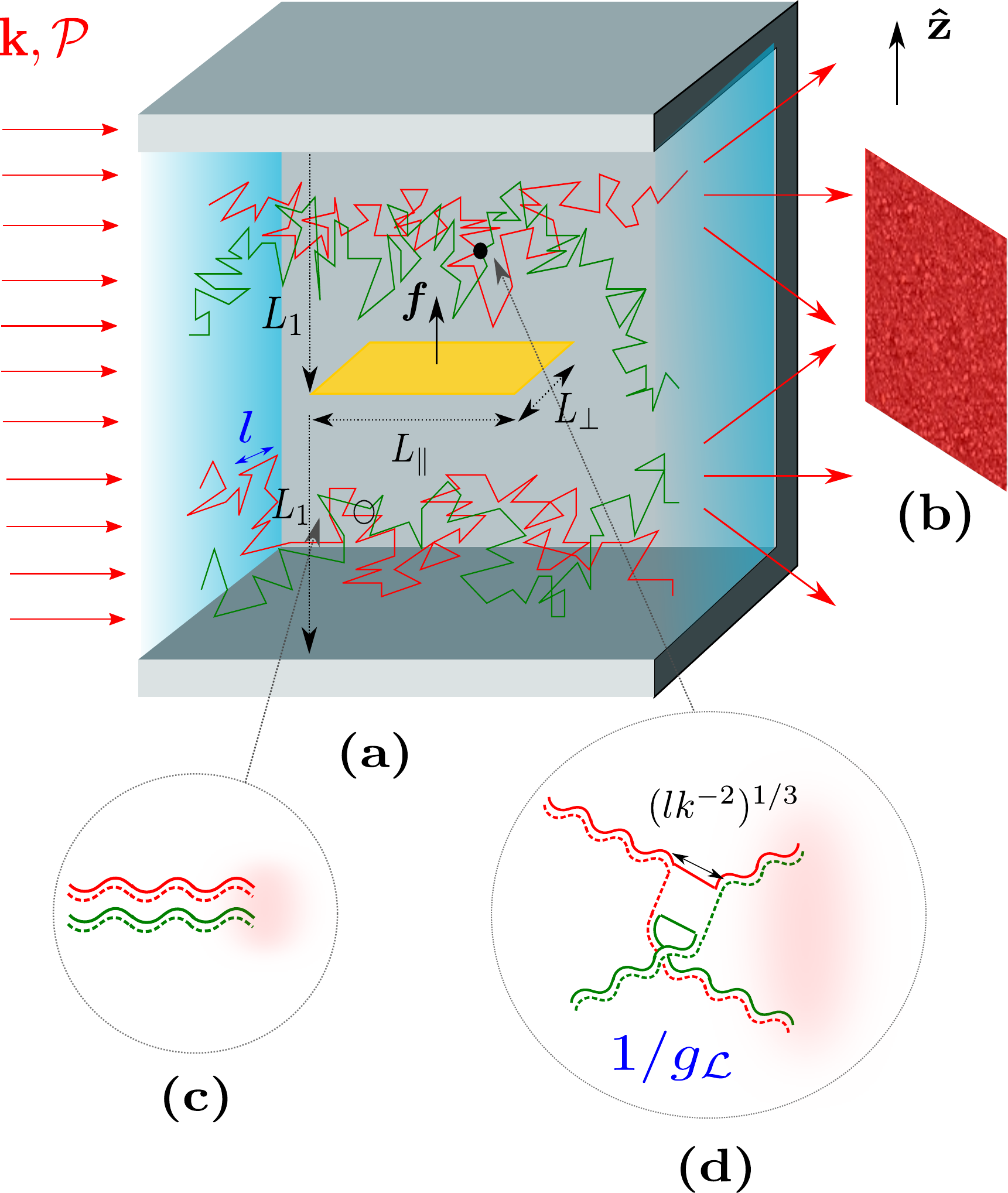}
\caption{ {\bf (a)} A monochromatic light beam of wave-number $k$ and power $\mathcal{P}$ experiences multiple elastic scattering in a random dielectric medium. For weak disorder, $k l \gg 1$, the average diffusive light intensity $I_D (\mathbf{r})$ is  represented by brownian-like trajectories. {\bf (b)} For each disorder realization, speckle patterns of bright and dark spots evidence spatial fluctuations of light intensity whose correlations are due to interference processes illustrated in {\bf (c)} and {\bf (d)}. {\bf (c)} Two phase-independent diffusive trajectories are built out of paired multiple scattering amplitudes -- solution of Eq.(\ref{helmoltz}) -- having opposite phases and pictured by two coupled (full and dotted) wave-shaped lines. These independent diffusive paths contribute to short range correlations.
{\bf (d)} Coherent long ranged correlations result from quantum crossings and a new pairing of phase-dependent amplitudes between two diffusive trajectories. The occurrence of a quantum crossing is proportional to  $1/g_{\mathcal{L}}$ (see text).
Coherent light fluctuations induce a fluctuating force $\bm{f}$ on a (suspended) plate immersed inside the scattering medium. When placed at equal distance $L_1$ from the lower and upper box edges, the average radiation force on both sides of the plate cancels out, leaving only the finite fluctuating part $\bm{f}$.}
\label{vis-abstract}
\end{figure}

In weakly disordered media, light intensity has long ranged spatial fluctuations (speckle) associated to mesoscopic coherent effects resulting from elastic multiple scattering. 
Here, we show that, unexpectedly, these intensity fluctuations lead to measurable FIF, $\bm{f} = \mathbf{f} - \langle\mathbf{f}\rangle$ (see Fig.\ref{vis-abstract}), on top of the disorder averaged radiation forces $\langle\mathbf{f}\rangle$ very similar in nature to Casimir forces.

The amplitude of the fluctuating radiation forces is 
\begin{equation}
\langle \bm{f}^2\rangle = \frac{1}{g_{\mathcal{L}}}\frac{\mathcal{P}^2}{v^2}\left(\mathcal{Q}_2 + \mathcal{Q_{\nu}}\right) \, .
\label{fif-form}
\end{equation}
This rather simple expression constitutes a central result of this work. It states that the fluctuating forces induced by coherent mesoscopic effects, besides their dependence upon the power $\mathcal{P}$ of the incoming light beam and the group velocity $v$, are driven by the dimensionless parameter  $g_{\mathcal{L}}$ which encapsulates both the geometry and the scattering properties of the random medium. It is the analog of conductance in electronic systems, henceforth called conductance. The two dimensionless numbers $\mathcal{Q}_2$ and $\mathcal{Q_{\nu}}$ depend on the shape of the system and on boundary conditions but not on its volume nor on scattering properties. These different quantities are detailed in the sequel.

Quite remarkably, spatially coherent light fluctuations can be thoroughly described using a Langevin equation, where a properly tailored noise accounts for mesoscopic coherent effects. This non intuitive result proves effective to establish Eq.(\ref{fif-form}), namely a relation between non equilibrium and Casimir physics on the one hand and coherent mesoscopic effects on the other hand. Moreover, this approach is of particular interest since it maps the problem of coherent multiple light scattering onto an effective non equilibrium light flow characterized by two parameters only, the diffusion coefficient $D$ and the strength of the noise $\sigma$, otherwise related by a Einstein relation.
The scarcity of measurable and temperature independent non equilibrium phenomena makes the present proposal particularly relevant to experimental inspections. Indeed, since light induced fluctuating forces depend on the easily tunable parameter $g_{\mathcal{L}}$, coherent multiple light scattering offers setups where FIF are significantly enhanced compared to other known situations \cite{Lamoreaux97,Lambrecht00,Munday09,Jourdan09, Mohideen98,Hertlein08}.

Consider a random and $d$-dimensional dielectric medium 
of volume $V = L^d$, illuminated by a monochromatic, scalar radiation \footnote{Polarization effects are usually decoupled from disorder. For more elaborations, see \cite{Akkermans}.} incident along the direction of unit vector $\mathbf{\widehat{k}}$ (see Fig.\ref{vis-abstract}.a). Inside the medium, the amplitude $E(\mathbf{r})$ of the radiation is solution of the scalar Helmholtz equation,
\begin{equation}
\Delta E(\mathbf{r})+k^2\left(1+\mu(\mathbf{r})\right)E(\mathbf{r})=s_0(\mathbf{r}) \, ,
\label{helmoltz}
\end{equation}
where $\mu(\mathbf{r})=\delta \epsilon(\mathbf{r})/\langle\epsilon\rangle$ denotes the fluctuation of the dielectric constant $\epsilon(\mathbf{r}) = \langle\epsilon\rangle+ \delta \epsilon(\mathbf{r})$ so that the wave number inside the medium is $k = \langle n \rangle \omega / v$ and $\langle n \rangle \equiv \sqrt{\langle \epsilon \rangle  / \epsilon_0}$ is the optical index averaged over disorder realizations  $\langle\cdot\cdot\cdot\rangle$. $s_0(\mathbf{r})$ is the source of the radiation. Disorder averaging allows to characterise the radiation propagation in the medium by the elastic mean free path $l$ defined as the average distance between two consecutive scattering events. \\
Multiple scattering solutions of the Helmholtz equation (\ref{helmoltz}) 
are notoriously difficult to obtain. In the weak disorder limit $kl\gg 1$, an equivalent description of the local radiation at a point $\mathbf{r}$ and propagating along a direction $\mathbf{\widehat{s}}$ is provided by the specific intensity $I(\mathbf{r,\widehat{s}})$, and the light current $\mathbf{j(r)} = v\overline{I(\mathbf{r,\widehat{s}})\, \mathbf{\widehat{s}}}$ averaged over all directions $\mathbf{\widehat{s}}$ \cite{Akkermans, Ishimaru} (see SM section 1.2). In this approach, the force exerted by light on an absorbing surface $S$ of normal vector $\mathbf{\widehat{n}}$, immersed inside the scattering medium, Fig.\ref{vis-abstract}.a, is \footnote{The specific intensity is directly related to the disorder averaged Poynting vector \cite{Ishimaru}. From this relation, we infer the absorbed energy $d {\mathcal E} = \mathbf{j(r)}\cdot \mathbf{\widehat{n}} \, dr dS / v^2$ and the mechanical radiation force $\mathbf{f} = - \boldsymbol{\nabla} {\mathcal E} |_{\mathbf{\widehat{n}}}  $ exerted along ${\mathbf{\widehat{n}}}$.  This expression of the radiation mechanical force is equivalent to the one derived from the Maxwell's stress tensor.}
\begin{equation}
\mathbf{f} = \frac{\mathbf{\widehat{n}}}{v^2}\int_{S}d\mathbf{r} \,  \,\mathbf{j(r)}\cdot\mathbf{\widehat{n}} \, \, .
\label{force_p1}
\end{equation}
A Fick's law of diffusion coefficient $D=vl /d$, 
\begin{equation}
\mathbf{j_D(r)}=-D \boldsymbol{\nabla} I_D(\mathbf{r}) \, ,
\label{fick}
\end{equation}
relates the disorder averaged light current $\mathbf{j_D(r)}$ to the disorder and direction averaged intensity $I_D(\mathbf{r})$. The latter obeys a diffusion equation, whose solutions have the generic form 
\begin{equation}
I_D(\mathbf{r}) = \frac{v \mathcal{P}}{D L}\, h(\mathbf{r})
\label{Id}
\end{equation}
where $h(\mathbf{r})$ is a dimensionless function determined by the geometry and boundary conditions and $L$ is a typical geometric size of the medium (see SM section 1.2). Inserting Eq.(\ref{fick}) into Eq.(\ref{force_p1}) allows to obtain the average radiation force $\langle\mathbf{f}\rangle$. Its value, for an incident light beam perpendicular to a surface  placed inside the medium at a distance $L$ from the incidence plane, is $\langle\mathbf{f}\rangle = \mathcal{P} T(L)/v$, where $T$ is the transmission coefficient (see SM section 2).


All phase dependent effects, responsible for speckle patterns (Fig.\ref{vis-abstract}.b), have been washed out in the disorder average diffusive limit underlying Eq.(\ref{fick}). A well defined semi-classical description enables to include coherent effects in a systematic way. The details of this generally cumbersome procedure are briefly sketched in the following paragraph and detailed in the SM. The main, first reading, message is that coherent effects lead to long range intensity fluctuations, whose  spatial correlations can be expanded as powers of $1/g_\mathcal{L}$, see Eq.(\ref{correlations}).

The semi-classical approach starts by noting (see Fig.\ref{vis-abstract}.c) that each  diffusive  trajectory is built from the pairing of two identical but time reversed multiple scattering amplitudes obtained from scattering solutions of Eq.(\ref{helmoltz}). By construction, these two amplitudes have opposite phases so that the resulting diffusive trajectory is phase independent. Unpairing these two sequences gives access to the underlying phase carried by each multiple scattering amplitude and thereby to phase coherent corrections. The aforementioned description makes profit of this remark to evaluate phase coherent corrections (see Fig.\ref{vis-abstract}.d). At a local crossing, two diffusive trajectories mutually exchange their phase so as to form two new phase independent diffusive trajectories. This local crossing -- or quantum crossing -- is a phase dependent correction propagated over long distances by means of diffusive trajectories \footnote{This phenomenological picture for coherent mesoscopic effects is presented at an introductory level in \cite{Akkermans}, section 1.7.}. The occurrence of a quantum crossing (Fig.\ref{vis-abstract}.d), in a disordered medium of volume $L^d$ is solely controlled by the conductance $g_{\mathcal{L}}$, a dimensionless parameter which depends on scattering properties and on the geometry of the medium. From now on and without loosing in generality, we consider the three dimensional ($d=3$) setup displayed in Fig.\ref{vis-abstract}. The conductance $g_{\mathcal{L}}$ is then of the form
\begin{equation}
g_{\mathcal{L}} \equiv \frac{k^2 l}{3\pi}\mathcal{L} \, 
\label{g}
\end{equation}
where the length $\mathcal{L}$ depends on the geometry (see later and SM section 5.2 for examples) \footnote{See chapter 12 in \cite{Akkermans}.}.
 In the weak disorder limit $k l \gg 1$, the conductance $g_{\mathcal{L}} \gg 1$ and small coherent corrections  generated by quantum crossings show up as powers of $1 / g_{\mathcal{L}}$. This scheme allows to expand spatial correlations of the fluctuating light intensity $\delta I (\mathbf{r}) \equiv I (\mathbf{r}) - I_D(\mathbf{r})$ as
\begin{equation}
\frac{\langle \delta I (\mathbf{r}) \delta I (\mathbf{r'})\rangle}{I_D(\mathbf{r})I_D(\mathbf{r'})} = C_1(\mathbf{r,r'})  + C_2(\mathbf{r,r'})  + C_3(\mathbf{r,r'}) \, \, .
\label{correlations}
\end{equation}
The first contribution $C_1(\mathbf{r,r'}) = \frac{2\pi l}{k^2}\delta(\mathbf{r-r'})$ (see Eq.(S47)) is short ranged and independent of $g_{\mathcal{L}}$. The two other contributions are long ranged, and respectively proportional to $1/g_{\mathcal{L}}$ and $1/ g_{\mathcal{L}}^2$. All three terms contribute to specific features of interference speckle patterns \cite{Goodman}, and have been measured in weakly disordered electronic and photonic media \cite{Akkermans,Kaveh87,Scheffold98,Scheffold97,Boer92,Stephen87}.

This $1/g_{\mathcal{L}}$ expansion can be obtained in a different but completely equivalent and elegant way by noting that quantum crossings occur at lengths of order $(lk^{-2})^{1/3} $, smaller than the elastic mean free path $l$. This allows to separate large scale $\left( \gg l \right)$ incoherent diffusive physics from small scale, coherent and phase preserving quantum crossings. This partition is described by a Langevin equation,
\begin{equation}
\mathbf{j(\mathbf{r})}=-D \boldsymbol{\nabla} I(\mathbf{r})+ \boldsymbol{\nu(r)} \, 
\label{langevin}
\end{equation}
which extends the Fick's law, Eq.(\ref{fick}), to the fluctuating, i.e. non  disorder averaged quantities $I (\mathbf{r}) \equiv I_D (\mathbf{r}) + \delta I (\mathbf{r})$ and $\mathbf{j(r)} \equiv \mathbf{j_D(r)} + \delta \mathbf{j(r)}$, by adding a zero average noise defined by the vector $\boldsymbol{\nu(r)}$. This picture, originally presented in \cite{Spivak87}, allows to reproduce the $1/g_{\mathcal{L}}$ expansion of Eq.(\ref{correlations}) by systematically including quantum crossings contributions into $\boldsymbol{\nu(r)}$. To lowest order in $1/g_{\mathcal{L}}$ (SM section 3), 
\begin{equation}
\langle\mathbf{\nu_{\alpha}(r)\nu_{\beta}(r')}\rangle=\delta_{\alpha \beta} \, c_0I_D^2(\mathbf{r})\, \delta(\mathbf{r}-\mathbf{r'}) 
\label{ki}
\end{equation}
where $c_0 \equiv \frac{2\pi l v^2}{3k^2}$.
We can rewrite the noise term under the form, $\boldsymbol{\nu(r)} =  \sqrt{\sigma}\, \boldsymbol{\eta}(\mathbf{\mathbf{r}})$, where $\langle\eta_{\alpha}(\mathbf{\mathbf{r}})\eta_{\beta}(\mathbf{\mathbf{r'}})\rangle = \delta_{\alpha\beta}\, \delta(\mathbf{r-r'})$ \footnote{The gaussian noise assumption in Eq.(\ref{ki}) is justified since the FIF do not depend on higher moments.}, with a strength, 
\begin{equation}
\sigma = c_0 \, I^2 _D(\mathbf{r}) \, ,
\label{sigma}
\end{equation}
which depends quadratically on the average diffusive radiation intensity $I_D(\mathbf{r})$ \footnote{The Langevin approach is valid for a weak noise, i.e. when $\sigma$ goes to zero with the system size. This requirement is here satisfied, see SM section 4.}.

This effective Langevin description, based on the two parameters $D$ and  $\sigma$, provides a complete hydrodynamic description of the  coherent light flow in the random medium. Moreover, it is appealing since its specific dependence upon a constant $D$ and a quadratic $\sigma$,  immediately draws a similarity with the  Kipnis-Marchioro-Presutti (KMP) process -- a heat transfer model for boundary driven one dimensional chains of mechanically uncoupled oscillators strongly out of equilibrium \cite{Kipnis82, Bertini05}, well described by the macroscopic fluctuation theory \cite{Bertini15}. A correspondence with this process is obtained by formally identifying the radiation intensity $I$ to the energy density, and $\mathbf{j}$ to the heat flow \footnote{The Langevin equation (\ref{langevin}) is time-independent, so that the correspondence is obtained by integrating the KMP Langevin equation over short time scales up to the elastic mean free time $\tau = l /c$ (see SM section 4).}. 
Despite this formal mapping, it is essential to note that the physical source of non equilibrium is very different in the two cases. While in the KMP model, energy density fluctuations result from thermal effects due to the coupling to two reservoirs at distinct temperatures, intensity fluctuations of the light flow result solely from the illumination of the random scattering medium.

A general Einstein relation exists which relates the  parameters $D$ and $\sigma$ characteristic of the hydrodynamic regime of strongly non equilibrium systems. It is given by $\sigma = D \chi(\mathbf{r})$, where $\chi(\mathbf{r})$ is the static compressibility \cite{Bertini05, Spohn}. For the coherent light flow,
\begin{equation}
\chi(\mathbf{r}) = 
\frac{c_0}{D}I^2_D(\mathbf{r}) \, ,
\label{Einstein}
\end{equation}
which from Eq.(\ref{sigma}), satisfies the Einstein relation (SM section 4). 



We are now in a position to calculate the radiation force $\mathbf{f}$,  which includes, on top of its average  $\langle\mathbf{f}\rangle$, a fluctuating (FIF) part $\bm{f} \equiv \mathbf{f} - \langle\mathbf{f}\rangle$ induced by intensity fluctuations. In the geometry of Fig.\ref{vis-abstract}, a dielectric plate, or membrane, of surface $S = L_\perp \times L_\parallel $, perpendicular to $\mathbf{\widehat{n}}= \mathbf{\widehat{z}}$, is inserted in the scattering medium so as to cancel by symmetry the average force $\langle\mathbf{f}\rangle$. The fluctuating part is readily obtained by substituting Eq.(\ref{langevin}) into Eq.(\ref{force_p1}) together with Eq.(\ref{fick}) and it is given by
\begin{align}
 \langle \bm{f}^2\rangle & = \frac{1}{v^4}\iint\limits_{S\times S}d\mathbf{r}d\mathbf{r'}[D^2\partial_z\partial_{z'}\langle \delta I(\mathbf{r})\delta I(\mathbf{r'})\rangle +\langle \nu_z(\mathbf{r})\nu_{z'}(\mathbf{r'})\rangle] \nonumber \\
 & \equiv \sum\limits_{j=1}^3 \bm{f_j}^2
 +\bm{f_{\nu}}^2 \, ,
\label{find}
\end{align}
where $\bm{f_j}^2 $ is the counterpart of the corresponding term in Eq.(\ref{correlations}) and $\bm{f_{\nu}}^2$ results from the noise term. The contribution $\bm{f_1}^2$ is always negligible compared to $\bm{f_{\nu}}^2$, as can easily be seen by considering the corresponding fluctuating forces on the faces of a cubic $L^3$ geometry with $L \gg l $ and without inner plate. The expression of $C_1$ together with Eqs.(\ref{Id},\ref{ki}), implies that $\bm{f_1}^2\sim \left(\frac{l}{L}\right)^2\bm{f_{\nu}}^2$, hence $\bm{f_1}^2$ is negligible. The term $\bm{f_3}^2$ induced by $C_3$ is of order $1/g_{\mathcal{L}}^2$ and therefore also negligible. Finally, the behaviour of $\bm{f_{\nu}}^2$ is readily obtained from Eqs.(\ref{Id},\ref{ki},\ref{Einstein}), namely $\bm{f_{\nu}}^2 =\frac{D}{v^4}\iint\limits_{S\times S}d\mathbf{r}d\mathbf{r'}\chi(\mathbf{r})\delta(\mathbf{r-r'}) =  \frac{1}{g_{\mathcal{L}}}\frac{\mathcal{P}^2}{v^2} \mathcal{Q}_\nu $, where $\mathcal{Q}_\nu$ is a dimensionless number characteristic of the system geometry. Then, as can be anticipated from Eq.(\ref{correlations}), $\bm{f_2}^2$ behaves like $1/g_{\mathcal{L}}$ and it is proportional to $\bm{f_{\nu}}^2$ (SM section 5.1), so that finally the fluctuating force has the general form Eq.(\ref{fif-form}) presented in the introductory paragraph.



\begin{figure}[t!]
\hspace*{-0.8cm}\includegraphics[width=1\columnwidth]{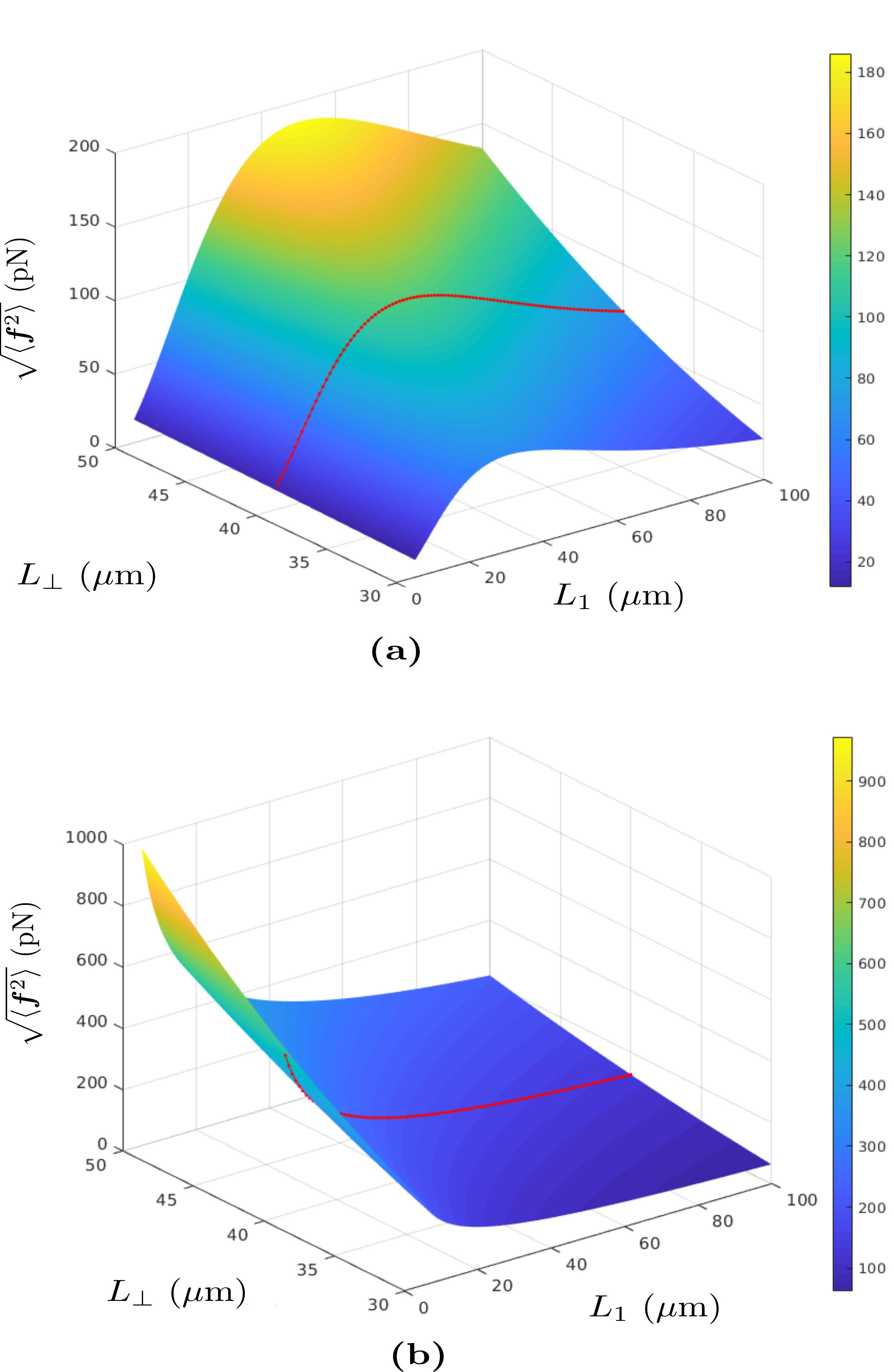}
\caption{Amplitude of $\sqrt{\langle \bm{f}^2\rangle} $ on the plate in Fig.\ref{vis-abstract} as a function of $L_1$ and $L_{\perp}$ with fixed $L_{\parallel}= 40\, \mu $m and $l = 1\,\mu $m. {\bf (a)} Absorbing plate with  $I_D(\mathbf{r})= 0$, so that $\langle\bm{f}^2\rangle = \bm{f}_2^2$. It vanishes in both  limits $L_1\to 0$ and $L_1\to +\infty$, which results from the form of $I_D(\mathbf{r})I_D(\mathbf{r'})C_2(\mathbf{r,r'})$ (see SM section 5.1 and 5.2). {\bf (b)} Reflecting plate where $\partial
_z I_D(\mathbf{r})= 0$, hence $\bm{f}_2 =0$ and $\langle\bm{f}^2\rangle =  \bm{f_{\nu}}^2$. From Eq.(\ref{ki}) and Eq.(\ref{find}), it appears that $\bm{f_{\nu}}^2$ scales like $1/\sqrt{L_1}$ (see SM section 5.2).  The red lines correspond to $L_{\perp} = 40\, \mu $m as in Table \ref{table-CL}.
}  
\label{tc2-pw-dir}
\end{figure}

We now evaluate more quantitatively the amplitude of the FIF in Eq.(\ref{fif-form}). Their dependence on the geometry and boundary conditions allows for a wide choice of parameters for control and amplification. Indeed, boundary conditions play an essential role in the determination of the dimensionless $\mathcal{Q}$'s, and even enable to measure independently $\bm{f}_2$ or $\bm{f_{\nu}}$ in Eq.(\ref{find}) (SM section 5.2). We highlight that a measurement of the sole contribution $\bm{f_{\nu}}$ of the noise induced by coherent effects, a by-product of our approach, cannot be achieved with other physical quantities, e.g. the transmission coefficient. Here, considering in the geometry of Fig.\ref{vis-abstract},
 an absorbing plate where  $I_D(\mathbf{r}) = 0$ (Fig.\ref{tc2-pw-dir}.a), selects only $\bm{f}_2$ which contributes with a maximum for an optimal value of $L_1$. Alternatively, inserting a reflective plate with $\partial_z I_D(\mathbf{r})=0$ selects $\bm{f_{\nu}}$ and leads to FIF with a power law dependence with $L_1$ (see Fig.\ref{tc2-pw-dir}.b and SM section 5.2). 
\begin{table}[t!]
\caption{Typical strength of light FIF in the setup of Fig.\ref{vis-abstract} obtained for visible light, $k \sim 10^7$ m$^{-1}$ and an elastic mean free path $l \simeq 1\, \mu $m i.e in a weakly disordered medium ($kl\sim 10$) and $v=2. 10^8$ m$\cdot$ s$^{-1}$. We consider the optimal case of reflecting cavity edges along $\mathbf{\widehat{x}}$ and absorbing edges along $\mathbf{\widehat{y}}$ (see text) and compare the cases of an absorbing and reflecting plate (Fig.\ref{tc2-pw-dir}). We obtain $g_{\mathcal{L}}=\frac{k^2l}{3\pi} \frac{L_1L_{\perp}L_{\parallel}}{\max(L_1^2,L_{\perp}^2,L_{\parallel}^2)}$ hence identifying the length $\mathcal{L}$ (see SM section 5.2). The amplitude of $\langle \bm{f}^2\rangle$ is calculated for different values of $L_1$ ranging from $5\, \mu$m to $100 \, \mu $m, with $L_{\perp}=L_{\parallel}=40 \, \mu $m, so that $L_1>l$ and $g_{\mathcal{L}}\gg 1$ in all cases. We choose $I = 10^9 \, $W$\cdot$ m$^{-2}$, an intensity strong enough to obtain measurable forces without altering  the medium.\label{table-CL}}
\medskip
\begin{tabular}{ccccc}
\toprule
 $\begin{array}{ll}
 \mbox{  }\\
 \mbox{ }
 \end{array}
 $  
& $L_1 (\mu$m$)$
& $\sqrt{\langle \bm{f}^2\rangle}$(pN)
& $\mathcal{Q}_2 + \mathcal{Q_{\nu}}$
& $g_{\mathcal{L}}$
\\
\hline
$ \begin{array}{ll}
    \mbox{Absorbing plate} \\
  \mathcal{Q}_\nu =0
\end{array}
$ 
& $\begin{array}{ll}
5\\
40\\
100
\end{array}$
& $\begin{array}{ll}
  13\\
  118\\
  68
\end{array}$
& $\begin{array}{ll}
  1.0\cdot 10^{-3}\\
  1.0\cdot 10^{-2}\\
  2.2\cdot 10^{-4}
\end{array}$ 
& $\begin{array}{ll}
  53\\
  424\\
  170
\end{array}$ 
\\
   \hline
$ \begin{array}{ll}
    \mbox{Reflecting plate} \\
  \mathcal{Q}_2 =0
\end{array}
$ 
& $\begin{array}{ll}
5\\
40\\
100
\end{array}$
&$\begin{array}{ll}
  567\\
  201\\
  127
\end{array}$ 
& $\begin{array}{ll}
  1.9\\
  2.3\cdot 10^{-2}\\
  7.6\cdot 10^{-4}
\end{array}$ 
& $\begin{array}{ll}
  53\\
  424\\
  170
\end{array}$ 
\\
\hline
\end{tabular} 
\end{table}

Sizeable efforts have been devoted to the development of high sensitivity cantilevers able to measure forces of weak amplitude \cite{Mohideen09}. We propose to observe mesoscopic FIF using an atomic force microscope, in a setup similar to \cite{Munday09} where Casimir-Lifshitz forces of a few piconewtons have been measured between a gold plate and a  gold coated sphere immersed in a liquid. 
Replacing the liquid by a weakly scattering medium $kl\sim 10$ and using square plates of size $40\, \mu $m$ \times 40\, \mu $m -- the typical size of the sphere used in \cite{Munday09} -- and illuminating the medium with a light beam of intensity $I\sim 10^9\,$W$\cdot$ $m^{-2}$, we expect light FIF of amplitude up to a few hundreds of piconewtons, i.e. strong enough to be detected. These  results are summarized in Table \ref{table-CL}.


 Eq.(\ref{fif-form}), together with the hydrodynamic description  of coherent effects based on the Langevin equation (\ref{langevin}), constitute the main results of this paper. Let us now discuss the scope of our findings in the context of ongoing research in mesoscopic physics and statistical mechanics, as well as applications. Aspects of diffusive light propagation, either incoherent or coherent, have already been thoroughly studied in the literature. For electronic quantum waves, the focus is mainly on transport properties, better accessible in mesoscopic devices and which stand as a favorite candidate to observe the elusive Anderson localization transition for large enough disorder. For radiation and other classical waves, transmission properties and long range correlations either spatial or spectral, have been also extensively studied. Despite these thorough investigations, mechanical effects resulting from coherent mesoscopic effects of diffusive light as presented here, have never been envisaged.  They open a new and alternative approach to the field. 
From a fundamental viewpoint, the existence of fluctuation induced forces easily and solely monitored by the dimensionless conductance Eq.(\ref{g}) has a threefold interest. First, the analogy here unveiled, between long range induced forces in a coherent mesoscopic light flow and in non equilibrium systems, should arouse experimental attention to observe such forces in the realm of radiation flow in Casimir physics. Second, coherent mechanical forces are sensitive to the disorder strength through the conductance $g_{\mathcal{L}}$. Hence, albeit non transport quantities, these forces can be used as a new effective probe to study the existence and criticality of Anderson localization transition both theoretically and experimentally. Third, potential applications of mechanical forces induced by a coherent diffusive radiation flow are diverse and promising: in addition to transmission measurements extensively used, they provide a new type of mechanical and sensitive sensors at submicronic scale rather easy to implement and useful in soft condensed matter, biophysics \cite{Bradonjic09}, nanoelectromechanical (NEMS) and quantum technologies \cite{Kippenberg07,Arcizet06}. Finally,  
we wish to highlight that the mapping we have presented between coherent light flow and out of equilibrium hydrodynamics is easily generalisable to other quantum or classical mesoscopic effects, e.g  in  nanoelectronics and superconductivity \cite{Henriet15}.
A clear asset of this type of approach is in its dependence upon two parameters only, thus making it a candidate to efficient machine learning algorithms.

This work was supported by the Israel Science Foundation Grant No.~924/09. We are grateful to Ohad Shpielberg for discussions, Boris Timchenko, Marc Soret and Igor Khmelnitsky for a critical reading of the manuscript and Yaroslav Don for his help in the preparation of the manuscript.


\bibliography{casimir}

\begin{thebibliography}{38}%
\makeatletter
\providecommand \@ifxundefined [1]{%
 \@ifx{#1\undefined}
}%
\providecommand \@ifnum [1]{%
 \ifnum #1\expandafter \@firstoftwo
 \else \expandafter \@secondoftwo
 \fi
}%
\providecommand \@ifx [1]{%
 \ifx #1\expandafter \@firstoftwo
 \else \expandafter \@secondoftwo
 \fi
}%
\providecommand \natexlab [1]{#1}%
\providecommand \enquote  [1]{``#1''}%
\providecommand \bibnamefont  [1]{#1}%
\providecommand \bibfnamefont [1]{#1}%
\providecommand \citenamefont [1]{#1}%
\providecommand \href@noop [0]{\@secondoftwo}%
\providecommand \href [0]{\begingroup \@sanitize@url \@href}%
\providecommand \@href[1]{\@@startlink{#1}\@@href}%
\providecommand \@@href[1]{\endgroup#1\@@endlink}%
\providecommand \@sanitize@url [0]{\catcode `\\12\catcode `\$12\catcode
  `\&12\catcode `\#12\catcode `\^12\catcode `\_12\catcode `\%12\relax}%
\providecommand \@@startlink[1]{}%
\providecommand \@@endlink[0]{}%
\providecommand \url  [0]{\begingroup\@sanitize@url \@url }%
\providecommand \@url [1]{\endgroup\@href {#1}{\urlprefix }}%
\providecommand \urlprefix  [0]{URL }%
\providecommand \Eprint [0]{\href }%
\providecommand \doibase [0]{http://dx.doi.org/}%
\providecommand \selectlanguage [0]{\@gobble}%
\providecommand \bibinfo  [0]{\@secondoftwo}%
\providecommand \bibfield  [0]{\@secondoftwo}%
\providecommand \translation [1]{[#1]}%
\providecommand \BibitemOpen [0]{}%
\providecommand \bibitemStop [0]{}%
\providecommand \bibitemNoStop [0]{.\EOS\space}%
\providecommand \EOS [0]{\spacefactor3000\relax}%
\providecommand \BibitemShut  [1]{\csname bibitem#1\endcsname}%
\let\auto@bib@innerbib\@empty
\bibitem [{\citenamefont {{M. Kardar, R. Golestanian}}(1999)}]{Kardar99}%
  \BibitemOpen
  \bibfield  {author} {\bibinfo {author} {\bibnamefont {{M. Kardar, R.
  Golestanian}}},\ }\bibfield  {title} {\enquote {\bibinfo {title} {{The
  Friction of Vacuum, and other Fluctuation-Induced Forces}},}\ }\href
  {\doibase 10.1103/RevModPhys.71.1233} {\bibfield  {journal} {\bibinfo
  {journal} {Rev. Mod. Phys.}\ }\textbf {\bibinfo {volume} {71}},\ \bibinfo
  {pages} {{}} (\bibinfo {year} {1999})}\BibitemShut {NoStop}%
\bibitem [{\citenamefont {{H. B. G. Casimir}}(1948)}]{Casimir48}%
  \BibitemOpen
  \bibfield  {author} {\bibinfo {author} {\bibnamefont {{H. B. G. Casimir}}},\
  }\bibfield  {title} {\enquote {\bibinfo {title} {{On the attraction between
  two perfectly conducting plates}},}\ }\href@noop {} {\bibfield  {journal}
  {\bibinfo  {journal} {Proc. Kon. Ned. Akad. Wet.}\ }\textbf {\bibinfo
  {volume} {B51}},\ \bibinfo {pages} {793--795} (\bibinfo {year}
  {1948})}\BibitemShut {NoStop}%
\bibitem [{\citenamefont {{M. E. Fisher, P.-G. de Gennes}}(1978)}]{DeGennes78}%
  \BibitemOpen
  \bibfield  {author} {\bibinfo {author} {\bibnamefont {{M. E. Fisher, P.-G. de
  Gennes}}},\ }\bibfield  {title} {\enquote {\bibinfo {title} {{Ph\'enom\`enes
  aux parois dans un m\'elange binaire critique}},}\ }\href@noop {} {\bibfield
  {journal} {\bibinfo  {journal} {C. R. Acad. Sci. Paris}\ }\textbf {\bibinfo
  {volume} {287}},\ \bibinfo {pages} {{}} (\bibinfo {year} {1978})}\BibitemShut
  {NoStop}%
\bibitem [{\citenamefont {{A. Aminov, Y. Kafri, M. Kardar}}(2015)}]{Kafri15}%
  \BibitemOpen
  \bibfield  {author} {\bibinfo {author} {\bibnamefont {{A. Aminov, Y. Kafri,
  M. Kardar}}},\ }\bibfield  {title} {\enquote {\bibinfo {title}
  {{Fluctuation-Induced Forces in Nonequilibrium Diffusive Dynamics}},}\ }\href
  {\doibase 10.1103/PhysRevLett.114.230602} {\bibfield  {journal} {\bibinfo
  {journal} {Phys. Rev. Lett.}\ }\textbf {\bibinfo {volume} {114}},\ \bibinfo
  {pages} {230602} (\bibinfo {year} {2015})}\BibitemShut {NoStop}%
\bibitem [{\citenamefont {{T. R. Kirkpatrick, J. M. Ortiz de Z\'arate, J. V.
  Sengers}}(2014)}]{Kirkpatrick14}%
  \BibitemOpen
  \bibfield  {author} {\bibinfo {author} {\bibnamefont {{T. R. Kirkpatrick, J.
  M. Ortiz de Z\'arate, J. V. Sengers}}},\ }\bibfield  {title} {\enquote
  {\bibinfo {title} {{Fluctuation-induced pressures in fluids in thermal
  nonequilibrium steady states}},}\ }\href {\doibase
  10.1103/PhysRevE.89.022145} {\bibfield  {journal} {\bibinfo  {journal} {Phys.
  Rev. E}\ }\textbf {\bibinfo {volume} {89}},\ \bibinfo {pages} {022145}
  (\bibinfo {year} {2014})}\BibitemShut {NoStop}%
\bibitem [{\citenamefont {{D. S. Dean, B.-S. Lu, A. C. Maggs, R.
  Podgornik}}(2016)}]{Dean16}%
  \BibitemOpen
  \bibfield  {author} {\bibinfo {author} {\bibnamefont {{D. S. Dean, B.-S. Lu,
  A. C. Maggs, R. Podgornik}}},\ }\bibfield  {title} {\enquote {\bibinfo
  {title} {{Nonequilibrium Tuning of the Thermal Casimir Effect}},}\ }\href
  {\doibase 10.1103/PhysRevLett.116.240602} {\bibfield  {journal} {\bibinfo
  {journal} {Phys. Rev. Lett.}\ }\textbf {\bibinfo {volume} {116}},\ \bibinfo
  {pages} {{}} (\bibinfo {year} {2016})}\BibitemShut {NoStop}%
\bibitem [{\citenamefont {{R. Messina, D. A. R. Dalvit, P. A. Maia Neto, A.
  Lambrecht and S. Reynaud}}(2009)}]{Reynaud09}%
  \BibitemOpen
  \bibfield  {author} {\bibinfo {author} {\bibnamefont {{R. Messina, D. A. R.
  Dalvit, P. A. Maia Neto, A. Lambrecht and S. Reynaud}}},\ }\bibfield  {title}
  {\enquote {\bibinfo {title} {{Dispersive interactions between atoms and
  nonplanar surfaces}},}\ }\href {\doibase 10.1103/PhysRevA.80.022119}
  {\bibfield  {journal} {\bibinfo  {journal} {Phys. Rev. A}\ }\textbf {\bibinfo
  {volume} {80}},\ \bibinfo {pages} {022119} (\bibinfo {year}
  {2009})}\BibitemShut {NoStop}%
\bibitem [{\citenamefont {Lamoreaux}(1997)}]{Lamoreaux97}%
  \BibitemOpen
  \bibfield  {author} {\bibinfo {author} {\bibfnamefont {S.~K.}\ \bibnamefont
  {Lamoreaux}},\ }\bibfield  {title} {\enquote {\bibinfo {title}
  {{Demonstration of the Casimir Force in the 0.6 to 6 $\mu$m Range}},}\ }\href
  {\doibase 10.1103/PhysRevLett.78.5} {\bibfield  {journal} {\bibinfo
  {journal} {Phys. Rev. Lett.}\ }\textbf {\bibinfo {volume} {78}},\ \bibinfo
  {pages} {{}} (\bibinfo {year} {1997})}\BibitemShut {NoStop}%
\bibitem [{\citenamefont {{A. Lambrecht, S. Reynaud}}(2000)}]{Lambrecht00}%
  \BibitemOpen
  \bibfield  {author} {\bibinfo {author} {\bibnamefont {{A. Lambrecht, S.
  Reynaud}}},\ }\bibfield  {title} {\enquote {\bibinfo {title} {{Casimir force
  between metallic mirrors}},}\ }\href {\doibase 10.1007/s100530050} {\bibfield
   {journal} {\bibinfo  {journal} {Eur. Phys. J. B}\ }\textbf {\bibinfo
  {volume} {8}},\ \bibinfo {pages} {309} (\bibinfo {year} {2000})}\BibitemShut
  {NoStop}%
\bibitem [{\citenamefont {{J. N. Munday, F. Capasso, V. A.
  Parsegian}}(2009)}]{Munday09}%
  \BibitemOpen
  \bibfield  {author} {\bibinfo {author} {\bibnamefont {{J. N. Munday, F.
  Capasso, V. A. Parsegian}}},\ }\bibfield  {title} {\enquote {\bibinfo {title}
  {{Measured long-range repulsive Casimir-Lifshitz forces}},}\ }\href {\doibase
  10.1038/nature07610} {\bibfield  {journal} {\bibinfo  {journal} {Nat.
  Letters.}\ }\textbf {\bibinfo {volume} {457}},\ \bibinfo {pages} {170--173}
  (\bibinfo {year} {2009})}\BibitemShut {NoStop}%
\bibitem [{\citenamefont {{G. Jourdan and A. Lambrecht and F. Comin and J.
  Chevrier}}(2009)}]{Jourdan09}%
  \BibitemOpen
  \bibfield  {author} {\bibinfo {author} {\bibnamefont {{G. Jourdan and A.
  Lambrecht and F. Comin and J. Chevrier}}},\ }\bibfield  {title} {\enquote
  {\bibinfo {title} {{Quantitative non-contact dynamic Casimir force
  measurements}},}\ }\href@noop {} {\bibfield  {journal} {\bibinfo  {journal}
  {Europhys. Lett.}\ }\textbf {\bibinfo {volume} {85}},\ \bibinfo {pages}
  {31001} (\bibinfo {year} {2009})}\BibitemShut {NoStop}%
\bibitem [{\citenamefont {Mohideen}\ and\ \citenamefont
  {Roy}(1998)}]{Mohideen98}%
  \BibitemOpen
  \bibfield  {author} {\bibinfo {author} {\bibfnamefont {U.}~\bibnamefont
  {Mohideen}}\ and\ \bibinfo {author} {\bibfnamefont {A.}~\bibnamefont {Roy}},\
  }\bibfield  {title} {\enquote {\bibinfo {title} {{Precision Measurement of
  the Casimir Force from 0.1 to 0.9 μm}},}\ }\href {\doibase
  10.1103/PhysRevLett.81.4549} {\bibfield  {journal} {\bibinfo  {journal}
  {Phys. Rev. Lett.}\ }\textbf {\bibinfo {volume} {81}},\ \bibinfo {pages}
  {4549} (\bibinfo {year} {1998})}\BibitemShut {NoStop}%
\bibitem [{\citenamefont {{C. Hertlein, L. Helden, A. Gambassi, S. Dietrich, C.
  Bechinger}}(2008)}]{Hertlein08}%
  \BibitemOpen
  \bibfield  {author} {\bibinfo {author} {\bibnamefont {{C. Hertlein, L.
  Helden, A. Gambassi, S. Dietrich, C. Bechinger}}},\ }\bibfield  {title}
  {\enquote {\bibinfo {title} {{Direct measurement of critical Casimir
  forces}},}\ }\href {\doibase 10.1038/nature06443} {\bibfield  {journal}
  {\bibinfo  {journal} {Nature}\ }\textbf {\bibinfo {volume} {451}},\ \bibinfo
  {pages} {172} (\bibinfo {year} {2008})}\BibitemShut {NoStop}%
\bibitem [{Note1()}]{Note1}%
  \BibitemOpen
  \bibinfo {note} {Polarization effects are usually decoupled from disorder.
  For more elaborations, see \cite {Akkermans}.}\BibitemShut {Stop}%
\bibitem [{\citenamefont {{E. Akkermans, G. Montambaux}}(2007)}]{Akkermans}%
  \BibitemOpen
  \bibfield  {author} {\bibinfo {author} {\bibnamefont {{E. Akkermans, G.
  Montambaux}}},\ }\href@noop {} {\emph {\bibinfo {title} {{Mesoscopic physics
  of electrons and photons}}}}\ (\bibinfo  {publisher} {Cambridge University
  Press},\ \bibinfo {year} {2007})\BibitemShut {NoStop}%
\bibitem [{\citenamefont {Ishimaru}(1978)}]{Ishimaru}%
  \BibitemOpen
  \bibfield  {author} {\bibinfo {author} {\bibfnamefont {A.}~\bibnamefont
  {Ishimaru}},\ }\href@noop {} {\emph {\bibinfo {title} {{Wave propagation and
  scattering in random media}}}}\ (\bibinfo  {publisher} {Academic Press},\
  \bibinfo {year} {1978})\BibitemShut {NoStop}%
\bibitem [{Note2()}]{Note2}%
  \BibitemOpen
  \bibinfo {note} {The specific intensity is directly related to the disorder
  averaged Poynting vector \cite {Ishimaru}. From this relation, we infer the
  absorbed energy $d {\protect \mathcal E} = \protect \mathbf {j(r)}\cdot
  \protect \mathbf {\setbox \z@ \hbox {\frozen@everymath \@emptytoks
  \mathsurround \z@ $\textstyle n$}\mathaccent "0362{n}} \protect \tmspace
  +\thinmuskip {.1667em} dr dS / v^2$ and the mechanical radiation force
  $\protect \mathbf {f} = - \protect \bm {\nabla } {\protect \mathcal E}
  |_{\protect \mathbf {\setbox \z@ \hbox {\frozen@everymath \@emptytoks
  \mathsurround \z@ $\textstyle n$}\mathaccent "0362{n}}} $ exerted along
  ${\protect \mathbf {\setbox \z@ \hbox {\frozen@everymath \@emptytoks
  \mathsurround \z@ $\textstyle n$}\mathaccent "0362{n}}}$. This expression of
  the radiation mechanical force is equivalent to the one derived from the
  Maxwell's stress tensor.}\BibitemShut {Stop}%
\bibitem [{Note3()}]{Note3}%
  \BibitemOpen
  \bibinfo {note} {This phenomenological picture for coherent mesoscopic
  effects is presented at an introductory level in \cite {Akkermans}, section
  1.7.}\BibitemShut {Stop}%
\bibitem [{Note4()}]{Note4}%
  \BibitemOpen
  \bibinfo {note} {See chapter 12 in \cite {Akkermans}.}\BibitemShut {Stop}%
\bibitem [{\citenamefont {Goodman}(2000)}]{Goodman}%
  \BibitemOpen
  \bibfield  {author} {\bibinfo {author} {\bibfnamefont {J.~W.}\ \bibnamefont
  {Goodman}},\ }\href@noop {} {\emph {\bibinfo {title} {{Statistical
  Optics}}}}\ (\bibinfo  {publisher} {Wiley Classics Library Edition},\
  \bibinfo {year} {2000})\BibitemShut {NoStop}%
\bibitem [{\citenamefont {{M. Kaveh, M. Rosenbluh, I.
  Freund}}(1987)}]{Kaveh87}%
  \BibitemOpen
  \bibfield  {author} {\bibinfo {author} {\bibnamefont {{M. Kaveh, M.
  Rosenbluh, I. Freund}}},\ }\bibfield  {title} {\enquote {\bibinfo {title}
  {{Speckle patterns permit direct observation of phase breaking}},}\ }\href
  {\doibase 10.1038/326778a0} {\bibfield  {journal} {\bibinfo  {journal}
  {Nature}\ }\textbf {\bibinfo {volume} {326}},\ \bibinfo {pages} {778--780}
  (\bibinfo {year} {1987})}\BibitemShut {NoStop}%
\bibitem [{\citenamefont {{F. Scheffold, G. Maret}}(1998)}]{Scheffold98}%
  \BibitemOpen
  \bibfield  {author} {\bibinfo {author} {\bibnamefont {{F. Scheffold, G.
  Maret}}},\ }\bibfield  {title} {\enquote {\bibinfo {title} {{Universal
  Conductance Fluctuations of Light}},}\ }\href {\doibase
  10.1103/PhysRevLett.81.5800} {\bibfield  {journal} {\bibinfo  {journal}
  {Phys. Rev. Lett.}\ }\textbf {\bibinfo {volume} {81}},\ \bibinfo {pages}
  {5800} (\bibinfo {year} {1998})}\BibitemShut {NoStop}%
\bibitem [{\citenamefont {{F. Scheffold, W. Hartl, G. Maret, E.
  Matijevic}}(1997)}]{Scheffold97}%
  \BibitemOpen
  \bibfield  {author} {\bibinfo {author} {\bibnamefont {{F. Scheffold, W.
  Hartl, G. Maret, E. Matijevic}}},\ }\bibfield  {title} {\enquote {\bibinfo
  {title} {{Observation of long-range correlations in temporal intensity
  fluctuations of light}},}\ }\href {\doibase 10.1103/PhysRevB.56.10942}
  {\bibfield  {journal} {\bibinfo  {journal} {Phys. Rev. B}\ }\textbf {\bibinfo
  {volume} {56}},\ \bibinfo {pages} {10942} (\bibinfo {year}
  {1997})}\BibitemShut {NoStop}%
\bibitem [{\citenamefont {{J. F. de Boer, M. P. van Albada, A.
  Lagendijk}}(1992)}]{Boer92}%
  \BibitemOpen
  \bibfield  {author} {\bibinfo {author} {\bibnamefont {{J. F. de Boer, M. P.
  van Albada, A. Lagendijk}}},\ }\bibfield  {title} {\enquote {\bibinfo {title}
  {{Transmission and intensity correlations in wave propagation through random
  media}},}\ }\href {\doibase 10.1103/PhysRevB.45.658} {\bibfield  {journal}
  {\bibinfo  {journal} {Phys. Rev. B}\ }\textbf {\bibinfo {volume} {45}},\
  \bibinfo {pages} {658} (\bibinfo {year} {1992})}\BibitemShut {NoStop}%
\bibitem [{\citenamefont {{M. J. Stephen, G. Cwilich}}(1987)}]{Stephen87}%
  \BibitemOpen
  \bibfield  {author} {\bibinfo {author} {\bibnamefont {{M. J. Stephen, G.
  Cwilich}}},\ }\bibfield  {title} {\enquote {\bibinfo {title} {{Intensity
  correlation functions and fluctuations in light scattered from a random
  medium}},}\ }\href {\doibase 10.1103/PhysRevLett.59.285} {\bibfield
  {journal} {\bibinfo  {journal} {Phys. Rev. Lett.}\ }\textbf {\bibinfo
  {volume} {59}},\ \bibinfo {pages} {285} (\bibinfo {year} {1987})}\BibitemShut
  {NoStop}%
\bibitem [{\citenamefont {{B. Z. Spivak, A. Yu. Zjuzin}}(1987)}]{Spivak87}%
  \BibitemOpen
  \bibfield  {author} {\bibinfo {author} {\bibnamefont {{B. Z. Spivak, A. Yu.
  Zjuzin}}},\ }\bibfield  {title} {\enquote {\bibinfo {title} {{Langevin
  description of mesoscopic fluctuations in random medium}},}\ }\href
  {http://jetp.ac.ru/cgi-bin/dn/e_066_03_0560} {\bibfield  {journal} {\bibinfo
  {journal} {Sov. Phys. JETP}\ }\textbf {\bibinfo {volume} {93}},\ \bibinfo
  {pages} {994--1006} (\bibinfo {year} {1987})}\BibitemShut {NoStop}%
\bibitem [{Note5()}]{Note5}%
  \BibitemOpen
  \bibinfo {note} {The gaussian noise assumption in Eq.(\ref {ki}) is justified
  since the FIF do not depend on higher moments.}\BibitemShut {Stop}%
\bibitem [{Note6()}]{Note6}%
  \BibitemOpen
  \bibinfo {note} {The Langevin approach is valid for a weak noise, i.e. when
  $\sigma $ goes to zero with the system size. This requirement is here
  satisfied, see SM section 4.}\BibitemShut {Stop}%
\bibitem [{\citenamefont {{C. Kipnis, C. Marchioro, E.
  Presutti}}(1982)}]{Kipnis82}%
  \BibitemOpen
  \bibfield  {author} {\bibinfo {author} {\bibnamefont {{C. Kipnis, C.
  Marchioro, E. Presutti}}},\ }\bibfield  {title} {\enquote {\bibinfo {title}
  {{Heat Flow in an Exactly Solvable Model}},}\ }\href {\doibase
  10.1007/BF01011740} {\bibfield  {journal} {\bibinfo  {journal} {J. Stat.
  Phys.}\ }\textbf {\bibinfo {volume} {27}},\ \bibinfo {pages} {65} (\bibinfo
  {year} {1982})}\BibitemShut {NoStop}%
\bibitem [{\citenamefont {{L. Bertini, D. Gabrielli, J. L.
  Lebowitz}}(2005)}]{Bertini05}%
  \BibitemOpen
  \bibfield  {author} {\bibinfo {author} {\bibnamefont {{L. Bertini, D.
  Gabrielli, J. L. Lebowitz}}},\ }\bibfield  {title} {\enquote {\bibinfo
  {title} {{Large Deviations for a Stochastic Model of Heat Flow}},}\ }\href
  {\doibase 10.1007/s10955-005-5527-2} {\bibfield  {journal} {\bibinfo
  {journal} {J. Stat. Phys.}\ }\textbf {\bibinfo {volume} {121}},\ \bibinfo
  {pages} {{}} (\bibinfo {year} {2005})}\BibitemShut {NoStop}%
\bibitem [{\citenamefont {{L. Bertini, A. De Sole, D. Gabrielli et
  al.}}(2015)}]{Bertini15}%
  \BibitemOpen
  \bibfield  {author} {\bibinfo {author} {\bibnamefont {{L. Bertini, A. De
  Sole, D. Gabrielli et al.}}},\ }\bibfield  {title} {\enquote {\bibinfo
  {title} {{Macroscopic fluctuation theory}},}\ }\href {\doibase
  10.1103/RevModPhys.87.593} {\bibfield  {journal} {\bibinfo  {journal} {Rev.
  Mod. Phys.}\ }\textbf {\bibinfo {volume} {87}},\ \bibinfo {pages} {593}
  (\bibinfo {year} {2015})}\BibitemShut {NoStop}%
\bibitem [{Note7()}]{Note7}%
  \BibitemOpen
  \bibinfo {note} {The Langevin equation (\ref {langevin}) is time-independent,
  so that the correspondence is obtained by integrating the KMP Langevin
  equation over short time scales up to the elastic mean free time $\tau = l
  /c$ (see SM section 4).}\BibitemShut {Stop}%
\bibitem [{\citenamefont {Spohn}(1991)}]{Spohn}%
  \BibitemOpen
  \bibfield  {author} {\bibinfo {author} {\bibfnamefont {H.}~\bibnamefont
  {Spohn}},\ }\href@noop {} {\emph {\bibinfo {title} {{Large Scale Dynamics of
  Interacting Particles}}}}\ (\bibinfo  {publisher} {Springer, Berlin},\
  \bibinfo {year} {1991})\BibitemShut {NoStop}%
\bibitem [{\citenamefont {R.~Castillo-Garza}\ and\ \citenamefont
  {Mohideen}(2009)}]{Mohideen09}%
  \BibitemOpen
  \bibfield  {author} {\bibinfo {author} {\bibfnamefont {D.~Yan}\ \bibnamefont
  {R.~Castillo-Garza}, \bibfnamefont {C.-C.~Chang}}\ and\ \bibinfo {author}
  {\bibfnamefont {U.}~\bibnamefont {Mohideen}},\ }\bibfield  {title} {\enquote
  {\bibinfo {title} {{Customized silicon cantilevers for Casimir force
  experiments using focused ion beam milling}},}\ }\href {\doibase
  10.1088/1742-6596/161/1/012005} {\bibfield  {journal} {\bibinfo  {journal}
  {J. Phys. Conf. Ser.}\ }\textbf {\bibinfo {volume} {161}},\ \bibinfo {pages}
  {012005} (\bibinfo {year} {2009})}\BibitemShut {NoStop}%
\bibitem [{\citenamefont {{K. Bradonji\'c, J. D. Swain, A. Widom, Y. N.
  Srivastava}}(2009)}]{Bradonjic09}%
  \BibitemOpen
  \bibfield  {author} {\bibinfo {author} {\bibnamefont {{K. Bradonji\'c, J. D.
  Swain, A. Widom, Y. N. Srivastava}}},\ }\bibfield  {title} {\enquote
  {\bibinfo {title} {{The Casimir Effect in Biology: The Role of Molecular
  Quantum Electrodynamics in Linear Aggregations of Red Blood Cells}},}\ }\href
  {\doibase 10.1088/1742-6596/161/1/012035} {\bibfield  {journal} {\bibinfo
  {journal} {J. Phys.: Conf. Ser}\ }\textbf {\bibinfo {volume} {161}},\
  \bibinfo {pages} {{}} (\bibinfo {year} {2009})}\BibitemShut {NoStop}%
\bibitem [{\citenamefont {{T. J. Kippenberg and K. J.
  Vahala}}(2007)}]{Kippenberg07}%
  \BibitemOpen
  \bibfield  {author} {\bibinfo {author} {\bibnamefont {{T. J. Kippenberg and
  K. J. Vahala}}},\ }\bibfield  {title} {\enquote {\bibinfo {title} {{Cavity
  Opto-Mechanics}},}\ }\href {\doibase 10.1364/OE.15.017172} {\bibfield
  {journal} {\bibinfo  {journal} {Opt. Express}\ }\textbf {\bibinfo {volume}
  {15}},\ \bibinfo {pages} {17172} (\bibinfo {year} {2007})}\BibitemShut
  {NoStop}%
\bibitem [{\citenamefont {{O. Arcizet, P.-F. Cohadon, T. Briant, M. Pinard, A.
  Heidmann}}(2006)}]{Arcizet06}%
  \BibitemOpen
  \bibfield  {author} {\bibinfo {author} {\bibnamefont {{O. Arcizet, P.-F.
  Cohadon, T. Briant, M. Pinard, A. Heidmann}}},\ }\bibfield  {title} {\enquote
  {\bibinfo {title} {{Radiation-pressure cooling and optomechanical instability
  of a micromirror}},}\ }\href {\doibase 10.1038/nature05244} {\bibfield
  {journal} {\bibinfo  {journal} {Nature}\ }\textbf {\bibinfo {volume} {444}},\
  \bibinfo {pages} {71} (\bibinfo {year} {2006})}\BibitemShut {NoStop}%
\bibitem [{\citenamefont {{L. Henriet, A. N. Jordan and K. Le
  Hur}}(2015)}]{Henriet15}%
  \BibitemOpen
  \bibfield  {author} {\bibinfo {author} {\bibnamefont {{L. Henriet, A. N.
  Jordan and K. Le Hur}}},\ }\bibfield  {title} {\enquote {\bibinfo {title}
  {{Electrical current from quantum vacuum fluctuations in nanoengines}},}\
  }\href {\doibase 10.1103/PhysRevB.92.125306} {\bibfield  {journal} {\bibinfo
  {journal} {Phys. Rev. B}\ }\textbf {\bibinfo {volume} {92}},\ \bibinfo
  {pages} {125306} (\bibinfo {year} {2015})}\BibitemShut {NoStop}%
\end{thebibliography}%


\begin{thebibliography}{7}
\providecommand{\natexlab}[1]{#1}
\providecommand{\url}[1]{\texttt{#1}}
\expandafter\ifx\csname urlstyle\endcsname\relax
  \providecommand{\doi}[1]{doi: #1}\else
  \providecommand{\doi}{doi: \begingroup \urlstyle{rm}\Url}\fi

\bibitem[{E. Akkermans, G. Montambaux}(2007)]{Akkermans}
{E. Akkermans, G. Montambaux}.
\newblock \emph{{"Mesoscopic physics of electrons and photons"}}.
\newblock Cambridge University Press, 2007.

\bibitem[Ishimaru(1978)]{Ishimaru}
A.~Ishimaru.
\newblock \emph{{"Wave propagation and scattering in random media"}}.
\newblock Academic Press, 1978.

\bibitem[{B. Z. Spivak, A. Yu. Zjuzin}(1987)]{Spivak87}
{B. Z. Spivak, A. Yu. Zjuzin}.
\newblock {"Langevin description of mesoscopic fluctuations in random medium"}.
\newblock \emph{Sov. Phys. JETP}, 93:\penalty0 994--1006, 1987.
\newblock URL \url{http://jetp.ac.ru/cgi-bin/dn/e_066_03_0560}.

\bibitem[{S. Hikami}(1981)]{Hikami81}
{S. Hikami}.
\newblock {"Anderson localization in a nonlinear-$\sigma$-model
  representation"}.
\newblock \emph{Phys. Rev. B}, 24:\penalty0 2671, 1981.
\newblock \doi{10.1103/PhysRevB.24.2671}.

\bibitem[{C. Kipnis, C. Marchioro, E. Presutti}(1982)]{Kipnis82}
{C. Kipnis, C. Marchioro, E. Presutti}.
\newblock {"Heat Flow in an Exactly Solvable Model"}.
\newblock \emph{J. Stat. Phys.}, 27:\penalty0 65, 1982.
\newblock \doi{10.1007/BF01011740}.

\bibitem[{L. Bertini, D. Gabrielli, J. L. Lebowitz}(2005)]{Bertini05}
{L. Bertini, D. Gabrielli, J. L. Lebowitz}.
\newblock {"Large Deviations for a Stochastic Model of Heat Flow"}.
\newblock \emph{J. Stat. Phys.}, 121, 2005.
\newblock \doi{10.1007/s10955-005-5527-2}.

\bibitem[Spohn(1991)]{Spohn}
H.~Spohn.
\newblock \emph{{"Large Scale Dynamics of Interacting Particles"}}.
\newblock Springer, Berlin, 1991.

\end{thebibliography}

\end{document}


\beginsupplement

\title{Supplementary material for \\
Fluctuating Forces Induced by Non Equilibrium and Coherent Light Flow}

\author{Ariane Soret$^{1,2}$, Karyn Le Hur$^2$, Eric Akkermans$^{1}$}
\affil{\textit{$^1$ Department of Physics, Technion -- Israel Institute of Technology, Haifa 3200003, Israel}}
\affil{\textit{$^2$Centre de Physique Th\'eorique, \'Ecole Polytechnique, CNRS, Universite Paris-Saclay, 91128 Palaiseau, France}}
\date{ }

\maketitle

\tableofcontents

\newpage

In this Supplementary Material, we provide more information on the notion of radiative forces in an elastic scattering medium. We also derive Eq.(8) from the general Eq.(6) and Eq.(7) in the Letter, and then we show that the light FIF take the universal form Eq.(12). Finally, we discuss quantitatively the magnitude of the FIF in simple cases.

\section{Classical light in scattering media}

In this section, we provide a derivation for the diffusion equation satisfied by the light intensity leading to the Fick's law in Eq.(3). 
We also discuss the relation with the Langevin equation (7). \\
Throughout this section, we consider a monochromatic scalar radiation, of wavenumber $k$, propagating through a scattering medium.
\subsection{Diffusion probability and structure factor}
\begin{figure}[h]
\centering
\includegraphics[scale = 0.2]{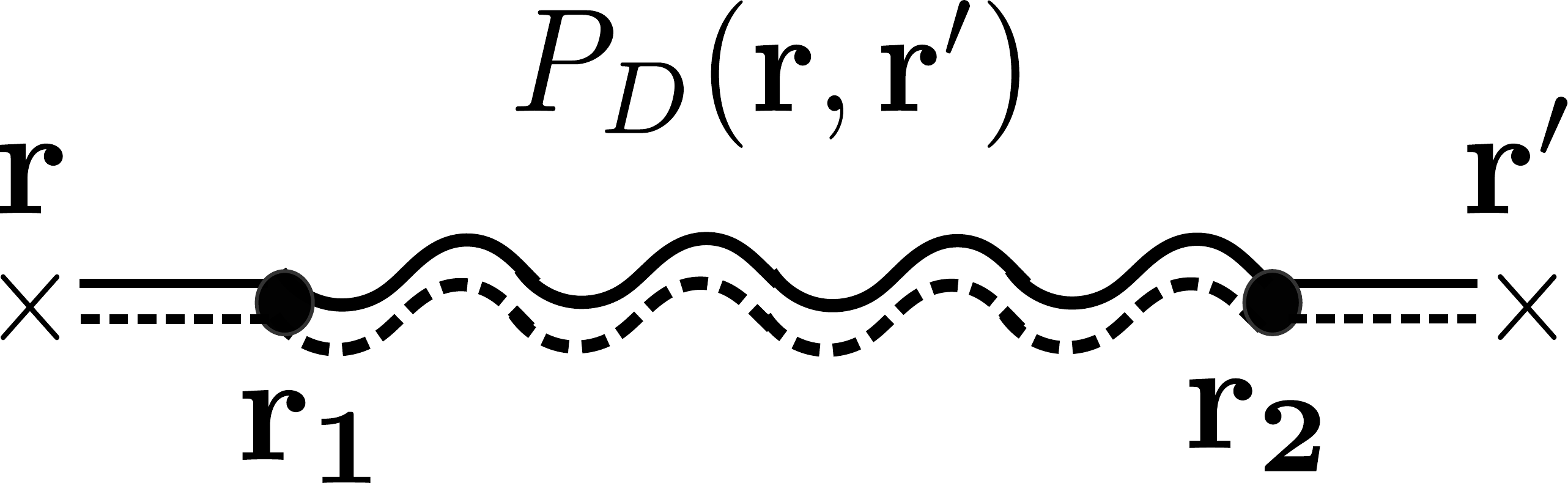}
\caption{Schematic representation of the diffusion probability $P_D(\mathbf{r,r'})$. The structure factor $\Gamma(\mathbf{r_1,r_2})$ is represented by the couple of wavelets between $\mathbf{r_1}$ and $\mathbf{r_2}$, while the couples of straight lines stand for the averaged Green's functions $|\langle G^R(\mathbf{r,r_1})\rangle|^2$ and $|\langle G^A(\mathbf{r',r_2})\rangle|^2$. }
\label{fig-pd}
\end{figure}

This section is a reminder of the known properties of light propagation in a scattering medium in the diffusion approximation, i.e. $kl\gg 1$ and after a large number of collisions.
Useful information on the monochromatic scalar radiation, of wavenumber $k$, is provided by the Green's function $G(\mathbf{r,r'})$ of the Helmoltz equation (1),
\begin{equation}
\left[ \Delta + k^2(1+\mu(\mathbf{r}))\right]G(\mathbf{r,r'}) = \delta(\mathbf{r-r'}) \, .
\label{helm}
\end{equation}
There are two solutions to Eq.(\ref{helm}), respectively the retarded and advanced Green's functions $G^R(\mathbf{r,r'})$, $G^A(\mathbf{r,r'})$. In vacuum, when $\mu(\mathbf{r}) = 0$, the solutions are
\begin{equation}
G_0^{R,A}(\mathbf{r,r'}) = \frac{e^{\pm ik|\mathbf{r-r'}|}}{4\pi|\mathbf{r-r'}|} \, .
\end{equation}
In a random scattering medium, $G^{R,A}(\mathbf{r,r'})$ take the form:
\begin{equation}
G^{R,A}(\mathbf{r,r'}) = \sum\limits_{N=1}^\infty\sum\limits_{\{\mathcal{C}_N\}} |\mathcal{A}(\mathbf{r,r',\mathcal{C_N}})|e^{\pm i k\mathcal{L}_N}
\label{green-function}
\end{equation}
where the double sum runs over all the possible sequences $\mathcal{C}_N$ of $N$ scatterers, with $N$ going from $1$ to infinity. $\mathcal{A}(\mathbf{r,r',\mathcal{C_N}})$ is the complex amplitude associated to the sequence of collisions $\mathcal{C_N}$, and $\mathcal{L}_N$ is the length of the corresponding scattering path.
From the Green's functions in Eq.(\ref{helm}), we define the probability $P(\mathbf{r,r'})$ for the radiation to scatter from $\mathbf{r}$ to $\mathbf{r'}$ by
\begin{equation}
P(\mathbf{r,r'}) = \frac{4\pi}{v}\langle G^{R}(\mathbf{r,r'}) G^{A}(\mathbf{r',r}) \rangle \, .
\label{prob-diff}
\end{equation}
The disorder averaged Green's functions can be calculated using a Dyson development (see section 3.2 in \cite{Akkermans}), which gives
\begin{equation}
\langle G^{R,A}(\mathbf{r,r'})\rangle = -\frac{1}{4\pi}\frac{e^{\pm ik|\mathbf{r-r'}|}}{|\mathbf{r-r'}|}e^{-|\mathbf{r-r'}|/2l},
\label{green-av}
\end{equation}
where the elastic mean free path $l$ is defined by: $\frac{4\pi}{l} = \langle V(\mathbf{q})V(\mathbf{q'})\rangle$, where $V(\mathbf{q})$ is the Fourier transform of the disorder potential $k^2\mu(\mathbf{r})$ in Eq.(\ref{helm}).
The disorder averaged product in Eq.(\ref{prob-diff}) is, however, difficult to derive exactly. Since averaging over disorder cancels out the terms in $G^{R}(\mathbf{r,r'}) G^{A}(\mathbf{r',r})$ which are taken over different sets of scatterers $(\mathbf{r_1,...,r_N})$, we keep only the terms with identical scattering sequences $\mathcal{C}_N$ -- which implies identical $\mathcal{L}_N$ and cancels the phase. We note $P_D(\mathbf{r,r'})$ the diffusion probability obtained from Eq.(\ref{prob-diff}) with this approximation. $P_D(\mathbf{r,r'})$ satisfies the equation
\begin{equation}
P_D(\mathbf{r,r'}) = \frac{4\pi}{v}\int\limits_V d\mathbf{r_1}d\mathbf{r_2}|\langle G^R(\mathbf{r,r_1})\rangle|^2 \Gamma(\mathbf{r_1,r_2})|\langle G^A(\mathbf{r',r_2})\rangle|^2
\label{pd}
\end{equation}
and is represented in Fig.\ref{fig-pd}.

The term $\Gamma(\mathbf{r_1,r_2})$ is the so called structure factor, and represents all the possible scattering sequences between $\mathbf{r_1}$ and $\mathbf{r_2}$. The collisions are assumed to be independent and $\Gamma(\mathbf{r_1,r_2})$ is therefore an infinite sum of collision sequences, which translates into an integral equation,
\begin{equation}
\Gamma(\mathbf{r_1,r_2}) = \frac{4\pi}{l}\delta(\mathbf{r_1-r_2}) + \frac{4\pi}{l}\int\limits_Vd\mathbf{r}\,\Gamma(\mathbf{r_1,r})|\langle G^R(\mathbf{r,r_2})\rangle|^2 \, .
\label{gamma}
\end{equation}
In the diffusive approximation, $\Gamma(\mathbf{r_1,r_2})$ varies slowly in space and we may use a Taylor expansion of $\Gamma(\mathbf{r_1,r_2})$ around $\mathbf{r_2}$. Keeping up to the quadratic term and substituing in Eq.(\ref{gamma}) leads to, after an integration by parts,
\begin{equation}
-D\Delta\Gamma(\mathbf{r_1,r_2}) = \frac{4\pi v}{l^2}\delta(\mathbf{r_1-r_2}) \, ,
\label{diff-gamma}
\end{equation}
where $D=\frac{vl}{3}$ is the diffusion coefficient. In the diffusive limit, we approximate $\Gamma(\mathbf{r_1,r_2})\simeq \Gamma(\mathbf{r,r'})$ in Eq.(\ref{pd}); using Eq.(\ref{green-av}), the Eq.(\ref{pd}) becomes
\begin{equation}
P_D(\mathbf{r,r'}) = \frac{l^2}{4\pi v}\Gamma(\mathbf{r,r'}) \, .
\label{gamma-pd}
\end{equation}
From Eq.(\ref{diff-gamma}) we deduce that $P_D(\mathbf{r,r'})$ satisfies
\begin{equation}
-D\Delta P_D(\mathbf{r,r'}) = \delta(\mathbf{r-r'}) \, .
\label{green-eq}
\end{equation}
The diffusion probability $P_D(\mathbf{r,r'})$ is hence the Green's function of a time independent diffusion equation, and has the generic form

\begin{equation}
P_d(\mathbf{r,r'}) = \frac{1}{DL}p(\mathbf{r,r'})
\label{pd-form}
\end{equation}
where $p(\mathbf{r,r'})$ is a dimensionless function depending on the boundary conditions.

\subsection{Fick's law and diffusion equation}
We now derive the Fick's law Eq.(3) and the diffusion equation for $I_D(\mathbf{r})$. In the scattering medium, the light intensity is related to the Green's functions in Eq.(\ref{green-function}) and the radiation source distribution $s_0(\mathbf{r})$ of the Helmoltz equation (1) by the Green's identity:
\begin{equation}
I(\mathbf{r}) = \iint\limits_{V\times V}d\mathbf{r_0}d\mathbf{r_0'}s_0(\mathbf{r_0})s_0(\mathbf{r_0'})G^R(\mathbf{r_0,r})G^A(\mathbf{r,r_0'}) \, .
\label{green-i}
\end{equation}
Note that the source term $s_0(\mathbf{r})$ is normalized so that $I(\mathbf{r})$ is related to the solution $E(\mathbf{r})$ of the Helmholtz equation (1) by $I(\mathbf{r}) = |E(\mathbf{r})|^2$. This normalization does not influence the derivation of Eq.(7-12) which constitute the main results of this work, and hence has no real importance here.
Calculating the above integral is a cumbersome task because of the complexity of the solutions $G^R(\mathbf{r,r'})$, $G^A(\mathbf{r,r'})$. An equivalent description of the radiation in the scattering medium is given by the specific intensity $I(\mathbf{\hat{s},r})$, whose behavior obeys the radiative transfer equation (see appendix A.5.2 in \cite{Akkermans}),
\begin{equation}
\mathbf{\hat{s}}\cdot\boldsymbol{\nabla} I(\mathbf{\hat{s},r}) = -\frac{1}{l}I(\mathbf{\hat{s},r})+\frac{1}{l}\overline{ I(\mathbf{\hat{s}',r})p(\mathbf{\hat{s}-\hat{s}'})}+\frac{1}{l}\gamma(\mathbf{\hat{s},r}) \, ,
\label{rad-transfer}
\end{equation} 
where the average is taken over all the directions $\mathbf{\hat{s}'}$ and with $\gamma(\mathbf{\hat{s},r})$ a light intensity source term placed inside the medium. Note that $\gamma(\mathbf{\hat{s},r})$ is different from $s_0(\mathbf{r})$, which is the source distribution for the electromagnetic radiation. The term $p(\mathbf{\hat{s}-\hat{s}'})$ accounts for scattering in other directions than $\mathbf{\hat{s}}$.
The disorder averaged intensity can be written as a sum of two contributions \cite{Akkermans,Ishimaru}: 
\begin{equation}
\langle I(\mathbf{\hat{s},r})\rangle = I_0(\mathbf{\hat{s},r})+I_D(\mathbf{\hat{s},r}) \, .
\label{i-sum}
\end{equation}
The first term $I_0(\mathbf{\hat{s},r})$ is the Drude-Boltzmann term and decreases exponentially due to scattering; it satisfies the equation
\begin{equation}
\mathbf{\hat{s}}\cdot\boldsymbol{\nabla} I_0(\mathbf{\hat{s},r}) =-\frac{1}{l}I_0(\mathbf{\hat{s},r}) \, .
\label{eq-diff-i0}
\end{equation}
The term $I_D(\mathbf{\hat{s},r})$ is the diffusive part, arising from multiple diffusion inside the disordered medium. We will further on neglect $I_0(\mathbf{\hat{s},r})$ in the expression of the average intensity: $\langle I(\mathbf{\hat{s},r})\rangle \simeq I_D(\mathbf{\hat{s},r})$. However, $I_0(\mathbf{\hat{s},r})$ does not completely disappear from the description since, as we will see, it behaves as a light source term for $I_D(\mathbf{\hat{s},r})$. Averaging Eq.(\ref{rad-transfer}) over disorder and using Eq.(\ref{eq-diff-i0}), we obtain, for $I_D(\mathbf{\hat{s},r})$,
\begin{equation}
\begin{array}{ll}
\mathbf{\hat{s}}\cdot\boldsymbol{\nabla} I_D(\mathbf{\hat{s},r})& = -\frac{1}{l}I_D(\mathbf{\hat{s},r})+\frac{1}{l}\overline{ I_D(\mathbf{\hat{s}',r})p(\mathbf{\hat{s}-\hat{s}'})}\\
\\
&+\frac{1}{l}\gamma_0(\mathbf{\hat{s},r})+\frac{1}{l}\gamma(\mathbf{\hat{s},r}) \, ,
\label{eq-ids}
\end{array}
\end{equation}
with $\gamma_0(\mathbf{\hat{s},r}) =\overline{ I_0(\mathbf{\hat{s}',r})p(\mathbf{\hat{s}-\hat{s}'})}$. We consider from now on that the scattering is isotropic, which implies that $p(\mathbf{\hat{s}-\hat{s}'})$ is independent of the angle between $\mathbf{\hat{s}}$ and $\mathbf{\hat{s}'}$. Under the additional assumption that the collisions with the scatterers are elastic (i.e. no absorption), $p(\mathbf{\hat{s}-\hat{s}'}) = 1$, and Eq.(\ref{eq-ids}) becomes
\begin{equation}
\mathbf{\hat{s}}\cdot\boldsymbol{\nabla} I_D(\mathbf{\hat{s},r}) = -\frac{1}{l}I_D(\mathbf{\hat{s},r})+\frac{1}{l} I_D(\mathbf{r})
+\frac{1}{l}I_0(\mathbf{r})+\frac{1}{l}\gamma(\mathbf{\hat{s},r}) \, .
\label{id-isotrope}
\end{equation}
Similarly to Eq.(\ref{i-sum}), we can write the average of the intensity current $\mathbf{j(r)}$ as a sum of two terms,
\begin{equation}
\langle\mathbf{j(r)}\rangle = \mathbf{j_0(r)}+\mathbf{j_D(\mathbf{r})} \, ,
\label{average-current}
\end{equation}
where $\mathbf{j_0}(\mathbf{r}) = v\overline{ I_0(\hat{s},\mathbf{r}) \hat{s}}$ is the current associated to the Drude-Boltzmann term, and $\mathbf{j_D}(\mathbf{r}) = v\overline{ I_D(\hat{s},\mathbf{r}) \hat{s}}$ the current associated to the diffusive contribution. In the main text, we neglected $\mathbf{j_0(r)}$ as it exponentially decreases to zero and does not play a role in the derivation of the FIF. We keep it here for a more complete description.
From Eq.(\ref{eq-diff-i0}), we obtain
\begin{equation}
\boldsymbol{\nabla}\cdot\mathbf{j_0(r)} = -\frac{v}{l}I_0(\mathbf{r})
\label{divj0}
\end{equation}
and, from Eq.(\ref{id-isotrope}), that the diffusive current satisfies the equation
\begin{equation}
\boldsymbol{\nabla}\cdot\mathbf{j_D(r)} = \frac{v}{l}I_0(\mathbf{r})+\frac{v}{l}\gamma(\mathbf{r})
\label{flux-cons}
\end{equation}
where $\gamma(\mathbf{r}) = \overline{\gamma(\mathbf{\hat{s},r})}$ is the light source distribution inside the medium. From Eq.(\ref{divj0}) and Eq.(\ref{flux-cons}), we obtain a continuity equation for the total average current,
\begin{equation}
\boldsymbol{\nabla}\cdot \langle\mathbf{j(r)}\rangle  = \frac{v}{l}\gamma(\mathbf{r}) \, .
\label{cons}
\end{equation}
In the case represented in Fig.1 and Fig.\ref{setup-sm}, the light source is located outside of the random medium, i.e. $\gamma(\mathbf{r})=0$ and the light flux is conserved.\\

In the diffusion approximation, $I_D(\mathbf{\hat{s},r})$ can be written 
\begin{equation}
I_D(\mathbf{\hat{s},r}) = I_D(\mathbf{r}) + \frac{3}{v}\mathbf{j_D(r)}\cdot\mathbf{\hat{s}} \, .
\label{id-approx}
\end{equation}
Replacing Eq.(\ref{id-approx}) in Eq.(\ref{id-isotrope}), we obtain
\begin{equation}
\mathbf{\hat{s}}\cdot\boldsymbol{\nabla}I_D(\mathbf{r}) + \frac{3}{v}\mathbf{\hat{s}}\cdot\boldsymbol{\nabla}(\mathbf{j_D(r)\cdot\hat{s}}) = -\frac{3}{vl}\mathbf{j_D(r)\cdot\hat{s}}+\frac{1}{l}\gamma(\mathbf{\hat{s},r})+\frac{1}{l}I_0(\mathbf{r}) \, .
\label{fick-aux}
\end{equation}
Projecting Eq.(\ref{fick-aux}) on $\mathbf{\hat{s}}$ and taking the average over all directions $\mathbf{\hat{s}}$ gives the Fick's law,
\begin{equation}
\mathbf{j_D(r)} = -D\boldsymbol{\nabla} I_D(\mathbf{r}) + v\overline{\mathbf{\hat{s}}\gamma(\mathbf{\hat{s},r})} \, .
\label{fick-gen}
\end{equation}
In the main text, we focus on the case where the medium is illuminated with an external light source, i.e. where no light source inside the medium: $\gamma(\mathbf{\hat{s},r}) = 0$. In this case, we recover the Fick's law Eq.(3).\\
Finally, combining Eq.(\ref{flux-cons}) and Eq.(\ref{fick-gen}), we obtain the diffusion equation satisfied by $I_D(\mathbf{r})$,
\begin{equation}
-D\Delta I_D(\mathbf{r}) = \frac{v}{l}I_0(\mathbf{r})+\frac{v}{l}\gamma(\mathbf{r})-v\overline{\mathbf{\hat{s}}\cdot\boldsymbol{\nabla}\gamma(\mathbf{r,\hat{s}})} \, .
\label{diff-eq}
\end{equation} 
Note that $I_0(\mathbf{r})$ plays the role of a source for the diffusive intensity $I_D(\mathbf{r})$; therefore, when the medium is illuminated by an external light source only, i.e. $\gamma = 0$, the right hand side of the diffusion equation Eq.(\ref{diff-eq}) is non zero.
\medskip
Using the Green's identity and the Green's function of the diffusion equation, $P_D(\mathbf{r,r'})$, introduced in Eq.(\ref{green-eq}), we obtain the expression of $I_D(\mathbf{r})$, 
\begin{equation}
I_D(\mathbf{r}) = \int\limits_V d\mathbf{r'}\gamma'(\mathbf{r'})P_D(\mathbf{r,r'}) \, ,
\label{green}
\end{equation} 
where $\gamma'(\mathbf{r}) = \frac{v}{l}I_0(\mathbf{r})+\frac{v}{l}\gamma(\mathbf{r})-v\overline{\mathbf{\hat{s}}\cdot\boldsymbol{\nabla}\gamma(\mathbf{r,\hat{s}})}$ is the source term in the diffusion equation (\ref{diff-eq}). We assume from now on that the light source is isotropic, which implies $\overline{\mathbf{\hat{s}}\cdot\boldsymbol{\nabla}\gamma(\mathbf{r,\hat{s}})}=0$ and
\begin{equation}
-D\Delta I_D(\mathbf{r}) = \frac{v}{l}I_0(\mathbf{r})+\frac{v}{l}\gamma(\mathbf{r}) \, .
\end{equation} 
%
\section{Average radiative force}
In this section, we derive the average force exerted by a monochromatic light beam on the boundaries of a scattering medium. We consider first the slab geometry represented in the Fig.1 and Fig.\ref{setup-sm}, and show that the average force of the surface $S_{\perp}=L_{\perp}(L_1+L_2)$ located at $y=L_{\parallel}$ is equal to $T(L_{\parallel})\mathcal{P}/v \, \mathbf{\hat{y}}$ where $T(L_{\parallel})$ is the transmission coefficient. We then consider the case of a closed surface, and deduce from Eq.(\ref{cons}) that the expression of the total radiative force on the surface is equal to $\mathcal{P}/v$, independently of disorder.

\subsection{Slab geometry with an external light source}

Consider a monochromatic plane wave of intensity $I$ emitted by an external light source and normally incident on a slab, of thickness $L_{\parallel}$, of an elastic scattering medium, as in Fig.\ref{setup-sm}. We derive the average force exerted by the scattered light on the opposite wall, of surface $S_{\perp}=L_{\perp}(L_1+L_2)$, located at $y=L_{\parallel}$. We follow the section A5.2.4 in \cite{Akkermans} for the derivation of the light current and intensity. By definition, the force on the surface $S_{\perp}$ is equal to
\begin{equation}
\langle\mathbf{f}\rangle = \frac{\mathbf{\hat{y}}}{v^2}\int\limits_{S_{\perp}}dxdz\,\langle j_y(x,y=L_{\parallel},z)\rangle \, .
\label{fslab}
\end{equation}
\\

We assume that $L_{\parallel}\ll L_{\perp},L_j$. In this slab geometry, the solution of Eq.(\ref{diff-eq}) with absorbing boundary conditions (see section 5.2 for details) is
\begin{equation}
I_D(\mathbf{r}) = 5I\frac{L_{\parallel}+l_0 - y}{L_{\parallel}+2l_0} - 3Ie^{-y/l} \, ,
\end{equation}
with $l_0 = \frac{2l}{3}$. From Fick's law Eq.(3), it follows that the diffusive current is equal to
\begin{equation}
\mathbf{j_D(r)} = 5ID\frac{1}{L_{\parallel}+2l_0} - 3I\frac{D}{l}e^{-y/l} \, .
\label{jdslab}
\end{equation}
The Drude intensity $I_0(\mathbf{r})$ is equal to
\begin{equation}
I_0(\mathbf{r}) = Ie^{-y/l} \, ,
\end{equation}
and the associated current, $\mathbf{j_0(r)}$, to
\begin{equation}
\mathbf{j_0(r)} = 3I\frac{D}{l}e^{-y/l} \, .
\label{j0slab}
\end{equation}
From Eq.(\ref{jdslab}) and Eq.(\ref{j0slab}), we deduce that the total current is constant and equal to
\begin{equation}
\langle\mathbf{j(r)}\rangle = 5ID\frac{1}{L_{\parallel}+2l_0} \, .
\label{jslab}
\end{equation}
Reinjecting Eq.(\ref{jslab}) in Eq.(\ref{fslab}), we obtain the force on the surface $S$:
\begin{equation}
\begin{array}{ll}
\langle\mathbf{f}\rangle = \frac{\mathbf{\hat{y}}}{v^2}S_{\perp}5ID\frac{1}{L_{\parallel}+2l_0}\\
\\
 = \frac{IS_{\perp}}{v}\frac{5l}{3(L_{\parallel}+2l_0)}\mathbf{\hat{y}}\\
 \\
 = \frac{\mathcal{P}}{v}T(L_{\parallel}) \mathbf{\hat{y}}
\end{array}
\end{equation}
where $\mathcal{P} = IS_{\perp}$ is the incoming light power, and $T(L_{\parallel}) = \frac{5l}{3(L_{\parallel}+2l_0)}$ the transmission coefficient.

\subsection{Closed surface}

Consider a volume $V$ delimited by a surface $\partial V$. We prove that the expression of the total average force on the closed surface $\partial V$ surrounding a light source of power $\mathcal{P}$ is equal to $\mathcal{P}/v$ regardless of the presence of disorder, with a group velocity $v$ which depends on the nature of the medium. The light source $\gamma$ inside the medium satisfies
\begin{equation}
\int\limits_Vd\mathbf{r}\gamma(\mathbf{r}) = l\mathcal{P} \, ,
\end{equation}
where $\mathcal{P}$ is the power of the light source. In the absence of scatterers, the energy flux is described by the Poynting vector $\bm{\Pi}(\mathbf{r})$. There is no energy dissipation, hence the flux of $\bm{\Pi}(\mathbf{r})$ through a closed surface $S$ around the source is independent of the size and shape of the surface. It is equal to the power of the source, $\oint_S d\mathbf{r}\, \bm{\Pi}(\mathbf{r})\cdot \mathbf{\hat{n}(r)} = \mathcal{P}$, with $\mathbf{\hat{n}(r)}$ the local unit vector normal to the surface at $\mathbf{r}$. The amplitude of the total normal radiative force in vacuum, $f_v$, is
\begin{equation}
f_{v} = \frac{1}{v}\oint\limits_S d\mathbf{r} \Pi(\mathbf{r})\cdot \mathbf{\hat{n}(r)} = \frac{\mathcal{P}}{v} \, .
\end{equation}
\medskip
In an elastic scattering medium, the total average force exerted by the scattered light on the boundary is by definition
\begin{equation}
\langle f \rangle  = \frac{1}{v^2}\oint\limits_{\partial V}\langle\mathbf{j(r)}\rangle\cdot\mathbf{\hat{n}} = \frac{1}{v^2}\int\limits_V \boldsymbol{\nabla}\cdot\langle\mathbf{j(r)}\rangle \, ,
\end{equation}
with $\langle\mathbf{j(r)}\rangle  = \mathbf{j_0(r)}+\mathbf{j_D(r)}$. Since $\boldsymbol{\nabla}\cdot\mathbf{j_0(r)} = -\frac{v}{l}I_0(\mathbf{r})$ and $\boldsymbol{\nabla}\cdot\mathbf{j_D(r)} = \frac{v}{l}I_0(\mathbf{r})+\gamma(\mathbf{r})$, we have
\begin{equation}
\begin{array}{ll}
\langle f \rangle & = \frac{1}{v^2}\int\limits_V -\frac{v}{l}I_0(\mathbf{r})+\frac{v}{l}I_0(\mathbf{r})+\frac{v}{l}\gamma(\mathbf{r})\\
\\
& = \frac{1}{cl}\int\limits_V \gamma(\mathbf{r})\\
 \\
& = \frac{\mathcal{P}}{v} \, ,
\end{array}
\end{equation}
which completes the proof.
%
\section{Derivation of the noise correlations}\label{der-current-corr}
We prove here the Eq.(8),
\begin{equation*}
\langle\mathbf{\nu_{\alpha}(r)\nu_{\beta}(r')}\rangle=\delta_{\alpha,\beta}c_0I_D^2(\mathbf{r})\delta(\mathbf{r}-\mathbf{r'}) \, ,
\end{equation*} 
with $c_0 = \frac{2\pi l v^2}{3k^2}$. This result was originally derived by \cite{Spivak87}, but we provide a more modern formulation and a more detailed derivation. The Eq.(8) is essential for the Langevin description since it translates mesoscopic interference processes in the Langevin language.

 The idea is to write an expression of the (connected) intensity correlation function $\langle\delta I(\mathbf{r})\delta I(\mathbf{r'})\rangle$, defined by 
\begin{equation}
\langle\delta I(\mathbf{r})\delta I(\mathbf{r'})\rangle = \langle I(\mathbf{r}) I(\mathbf{r'})\rangle - I_D(\mathbf{r})I_D(\mathbf{r'}),
\end{equation}
using, on the one hand, the noise correlations $\langle\nu_{\alpha}(\mathbf{r})\nu_{\beta}(\mathbf{r'})\rangle$, and on the other hand, a diagrammatic description.
From the Langevin equation Eq.(7), the continuity equation $\boldsymbol{\nabla}\cdot\mathbf{j(r)} = 0$ and Fick's law Eq.(3), we deduce a Langevin equation for the intensity fluctuation $\delta I(\mathbf{r})$, completed with a continuity equation,
\begin{equation}
\left\{
\begin{array}{ll}
\delta \mathbf{j(r)} = -D\boldsymbol{\nabla} \delta I(\mathbf{r})+\bm{\nu(r)}\\
\\
\boldsymbol{\nabla}\cdot \delta \mathbf{j(r)} = 0
\end{array}
\right.
\label{syst}
\end{equation}
From Eqs.(\ref{syst}), using an integration by parts, we obtain
\begin{equation}
\langle\delta I(\mathbf{r})\delta I(\mathbf{r'})\rangle = \frac{\sigma}{v^2}\delta(\mathbf{r-r'})+C(\mathbf{r,r'}) \, ,
\label{corr-source-tot}
\end{equation}
with $C(\mathbf{r,r'})$ a long-ranged function, equal to
\begin{equation}
\!\!\! C(\mathbf{r,r'})=
\iint\limits_{V\times V}d\mathbf{r_1}d\mathbf{r_2}
\boldsymbol{\nabla}_{1,\alpha}P_D(\mathbf{r,r_1})\cdot\boldsymbol{\nabla}_{2,\beta}P_D(\mathbf{r',r_2})\langle\nu_{\alpha}(\mathbf{r_1})\nu_{\beta}(\mathbf{r_2})\rangle \, .
\label{corr-source}
\end{equation}
%
On the other hand, $\langle\delta I(\mathbf{r})\delta I(\mathbf{r'})\rangle$ can be derived from Eq.(\ref{green-i}) using a perturbative development of products of the Green's functions. In the main text, Eq.(6), we kept the first three terms of the development,
\begin{equation}
\langle\delta I(\mathbf{r})\delta I(\mathbf{r'})\rangle = I_D(\mathbf{r})I_D(\mathbf{r'})\left[C_1(\mathbf{r,r'})+C_2(\mathbf{r,r'})+C_3(\mathbf{r,r'})\right] \, ,
\end{equation}
the sum of a short ranged function and of two long ranged functions. By identification with Eq.(\ref{corr-source-tot}), we obtain the identity 
\begin{equation}
C(\mathbf{r,r'}) = I_D(\mathbf{r})I_D(\mathbf{r'})\left[C_2(\mathbf{r,r'})+C_3(\mathbf{r,r'})\right]
\end{equation}
A scaling argument using Eqs.(\ref{pd-form}, \ref{corr-source}, 4) leads to a generic expression for $\langle\nu_{\alpha}(\mathbf{r})\nu_{\beta}(\mathbf{r'})\rangle$,
\begin{equation}
\langle\nu_{\alpha}(\mathbf{r})\nu_{\beta}(\mathbf{r'})\rangle = \frac{D^2}{g_{\mathcal{L}}}\mathcal{F}_2(I_d) +  \frac{D^2}{g_{\mathcal{L}}^2}\mathcal{F}_3(I_d)
\end{equation}
where $\mathcal{F}_j$ are functionals of $I_d$.

\begin{figure*}
\includegraphics[scale=0.3]{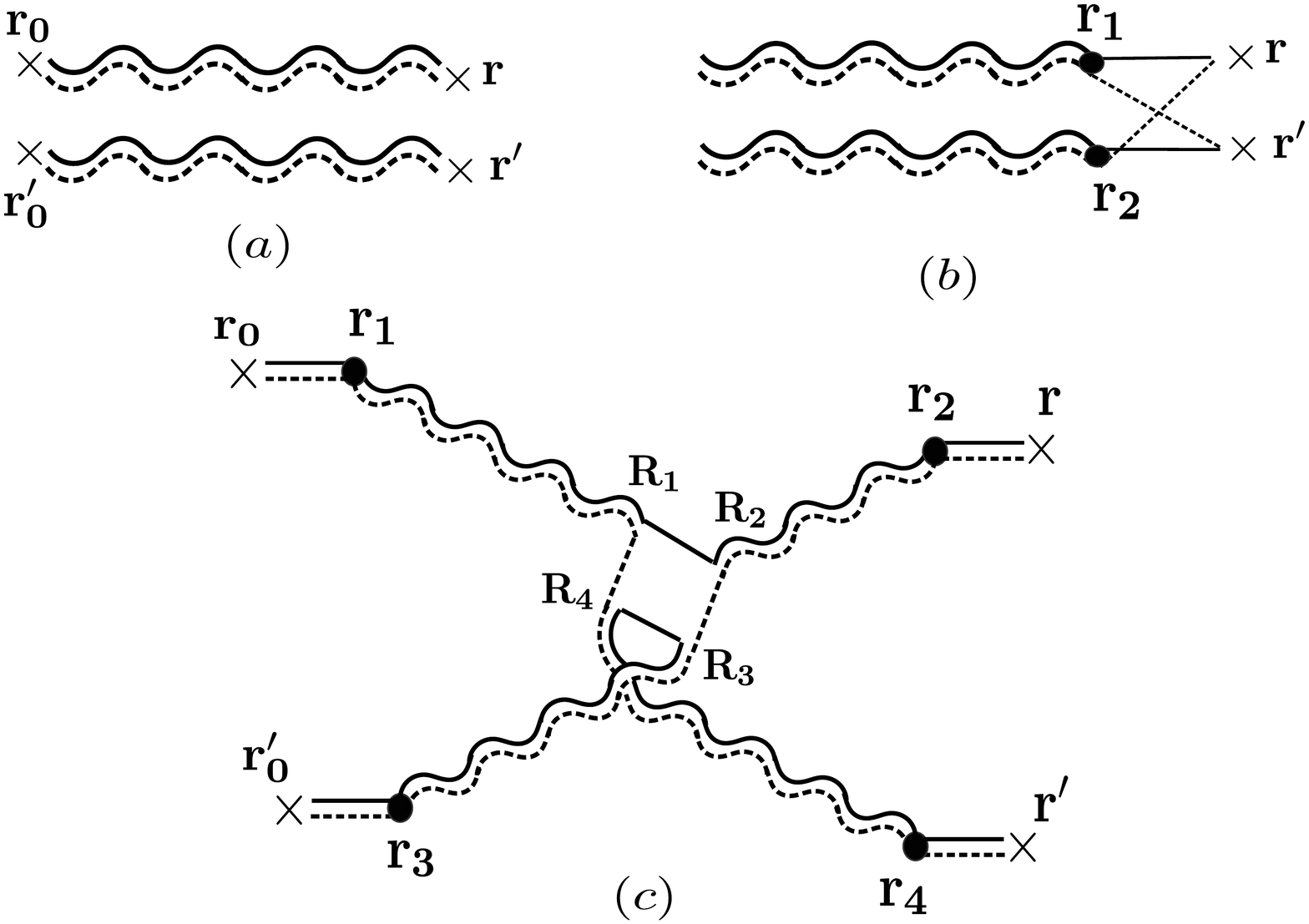}
\caption{(a) Two diffusion paths propagating without interacting. (b) Representation of the short-ranged term in the correlation function of the intensity fluctuations. (c) Diagram corresponding to the $C_2$ term of the intensity correlations. The square represents a Hikami box, describing one quantum crossing between two diffusion paths.}
\label{corr-meso}
\end{figure*}

Since we seek the main contribution to the noise correlation function, we neglect the term $C_3$ and we rewrite the correlation function in the form
\begin{equation}
\langle\delta I(\mathbf{r})\delta I(\mathbf{r'})\rangle = \langle\delta I(\mathbf{r})\delta I(\mathbf{r'})\rangle^{(1)}+\langle\delta I(\mathbf{r})\delta I(\mathbf{r'})\rangle^{(2)} \, ,
\end{equation} 
where $\langle\delta I(\mathbf{r})\delta I(\mathbf{r'})\rangle^{(j)} = I_D(\mathbf{r})I_D(\mathbf{r'})C_j(\mathbf{r,r'})$. The short ranged term $\langle\delta I(\mathbf{r})\delta I(\mathbf{r'})\rangle^{(1)}$ corresponds to the diagram Fig.\ref{corr-meso}.b, where two wave packets propagate by multiple scattering and cross between $\mathbf{r_1}$, $\mathbf{r_2}$ and $\mathbf{r}$, $\mathbf{r'}$. It is equal to
\begin{equation}
\langle\delta I(\mathbf{r})\delta I(\mathbf{r'})\rangle^{(1)} = \frac{2\pi l}{k^2}I_D(\mathbf{r})^2\delta(\mathbf{r-r'}) \, .
\label{c1}
\end{equation}
The long ranged term, represented on Fig.\ref{corr-meso}.c, is equal to
\begin{equation}
\begin{array}{ll}
\langle \delta I(\mathbf{r})\delta I(\mathbf{r'})\rangle^{(2)}  =
\left(\frac{4\pi}{v}\right)^2\int\limits_{V}d\mathbf{r_0}d\mathbf{r_0'}\gamma'(\mathbf{r_0})\gamma'(\mathbf{r_0'}) \int\limits_{V} \prod\limits_{i=1}^4 d\mathbf{R_i}\prod\limits_{j=1}^4 d\mathbf{r_j}\\
\\
|G^R(\mathbf{r_0,r_1})|^2|G^R(\mathbf{r_0',r_3})|^2 H(\mathbf{R_i}) \Gamma(\mathbf{r_1,R_1})\Gamma(\mathbf{r_3,R_3})\\
\\
\times\Gamma(\mathbf{R_2,r_2})\Gamma(\mathbf{R_4,r_4})|G^R(\mathbf{r_2,r})|^2|G^R(\mathbf{r_4,r'})|^2 \, ,
\end{array}
\label{c2-diag}
\end{equation}
with $\gamma'(\mathbf{r})$ the source term for $I_D(\mathbf{r})$ introduced in Eq.(\ref{green}), and where $H(\mathbf{R_i})$ is a so called Hikami box \cite{Hikami81} - an operator which precisely describes an interaction between two diffusion paths. Its analytical expression is 
\begin{equation}
H(\mathbf{R_i}) = 2h_4\int\limits_V d\mathbf{R}\prod_{i=1}^4\delta(\mathbf{R-R_i})\boldsymbol{\nabla}_{\mathbf{R_2}}\cdot\boldsymbol{\nabla}_{\mathbf{R_4}} \, ,
\end{equation}
with $h_4 = \frac{l^5}{48\pi k^2}$, as detailed in \cite{Akkermans}. 
In the diffusive limit, $\Gamma(\mathbf{r,r'})$ varies slowly and we can replace $\Gamma(\mathbf{r_1,R_1})\simeq\Gamma(\mathbf{r_0,R_1})$, $\Gamma(\mathbf{r_3,R_3})\simeq\Gamma(\mathbf{r_0',R_3})$, $\Gamma(\mathbf{R_2,r_2})\simeq\Gamma(\mathbf{R_2,r})$ and $\Gamma(\mathbf{R_4,r_4})\simeq\Gamma(\mathbf{R_4,r'})$ in Eq.(\ref{c2-diag}). The averaged Green's function are then decoupled; their integral is equal to $\frac{l}{4\pi}$. Applying the Hikami box operator on the functions $\Gamma$, we obtain,
\begin{equation}
\begin{array}{ll}
\langle \delta I(\mathbf{r})\delta I(\mathbf{r'})\rangle^{(2)}  = 2h_4\left(\frac{4\pi}{v}\right)^2\left(\frac{l}{4\pi}\right)^4\int\limits_{V}d\mathbf{r_0}d\mathbf{r_0'}\gamma'(\mathbf{r_0})\gamma'(\mathbf{r_0'})\\
\\
\int\limits_Vd\mathbf{R}\int\limits_V\prod\limits_{i=1}^4d\mathbf{R_i}\delta(\mathbf{R-R_i})\boldsymbol{\nabla}_{\mathbf{R_2}}\cdot\boldsymbol{\nabla}_{\mathbf{R_4}}\left[\Gamma(\mathbf{r_0,R_1})\Gamma(\mathbf{r_0',R_3})\Gamma(\mathbf{R_2,r})\Gamma(\mathbf{R_4,r'})\right]\\
\\
= 2h_4\left(\frac{4\pi}{v}\right)^2\left(\frac{l}{4\pi}\right)^4\\
\\
\times\int\limits_{V}d\mathbf{r_0}d\mathbf{r_0'}\gamma'(\mathbf{r_0})\gamma'(\mathbf{r_0'})\int\limits_Vd\mathbf{R}\, \Gamma(\mathbf{r_0,R})\Gamma(\mathbf{r_0',R})\boldsymbol{\nabla}_{\mathbf{R}}\Gamma(\mathbf{R,r})\cdot\boldsymbol{\nabla}_{\mathbf{R}}\Gamma(\mathbf{R,r'}) \, .
\end{array}
\end{equation}
Using Eq.(\ref{gamma-pd}) and Eq.(\ref{green}), we obtain finally
\begin{equation}
\langle\delta I(\mathbf{r})\delta I(\mathbf{r'})\rangle^{(2)} 
= c_0\int\limits_V d\mathbf{R}\,I_D^2(\mathbf{R})\boldsymbol{\nabla}_{\mathbf{R}}P_D(\mathbf{R,r})\cdot\boldsymbol{\nabla}_{\mathbf{R}}P_D(\mathbf{R,r'}) \, ,
\label{corrI2}
\end{equation}
with $c_0 = \frac{2\pi l v^2}{3k^2}$.
Comparing Eq.(\ref{corrI2}) with Eq.(\ref{corr-source}) allows to identify the noise correlation function,
\begin{equation*}
\langle\nu_{\alpha}(\mathbf{r})\nu_{\beta}(\mathbf{r'})\rangle = \delta_{\alpha,\beta}\frac{2\pi lv^2}{3k^2}I_D^2(\mathbf{r})\delta(\mathbf{r}-\mathbf{r'}) \, ,
\end{equation*} 
which completes the proof of Eq.(8).

As we highlighted in the main text, the amplitude of the noise depends on the average value $I_D(\mathbf{r})$, which is different from usual stochastic processes. Furthermore, the term $\sqrt{\frac{2\pi lv^2}{3k^2}}$ in the amplitude of the noise comes directly from the amplitude of the Hikami box and of the value of the other quantities involved in the crossing of two diffusion paths - the structure factor and the average Green's functions $G^{R,A}$. The factor $\sqrt{\frac{2\pi lv^2}{3k^2}}$ is the signature of the mesoscopic interference processes which lead to long ranged intensity correlations.
Note that we restricted ourselves to the derivation of the main term of the noise correlation function. We can use the same reasoning to derive the next term of the noise correlation function which, once reinjected in Eq.(\ref{corr-source}), gives the term $\langle \delta I(\mathbf{r})\delta I(\mathbf{r'})\rangle^{(3)} = I_D(\mathbf{r})I_D(\mathbf{r'})C_3(\mathbf{r,r'})$. Since $C_3$ has a negligible amplitude compared to $C_2$, so do the corresponding noise correlation terms. 
%

\section{Coherent light flow in the macroscopic fluctuation theory approach}
We derive here the compressibility Eq.(10) and establish the correspondence between coherent light and the KMP model. Finally, we discuss the validity of the Langevin approach.

\medskip

The KMP model describes a time dependent heat transfer process on a one-dimensional lattice. A chain of mechanically uncoupled oscillators is coupled at its extremities to two thermal baths at different temperatures, which induces a heat flow governed by nonharmonic effects on the lattice. These effects are accounted by an effective stochastic process which redistributes the energy between nearest neighbors \cite{Kipnis82}. This process can be described macroscopically by means of a Langevin equation completed by a continuity equation,
\begin{equation}
\left\{
\begin{array}{ll}
\mathbf{J}(\mathbf{r},t) = -D\nabla\rho(\mathbf{r},t) +\sqrt{\sigma}\bm{\eta}(\mathbf{r},t)\\
\\
\partial_t \rho(\mathbf{r},t) = -\nabla\cdot\mathbf{J}(\mathbf{r},t)
\end{array}
\right.
\label{langevin}
\end{equation}
with $\rho(\mathbf{r},t)$ the energy density, $\mathbf{J}(\mathbf{r},t)$ the thermal flux and $\bm{\eta}(\mathbf{r},t)$ a Gaussian noise,
\begin{equation}
\left\{
\begin{array}{ll}
\langle\bm{\eta}(\mathbf{r},t)\rangle = 0\\
\\
\langle\eta_{\alpha}(\mathbf{r},t)\eta_{\beta}(\mathbf{r'},t')\rangle = \delta_{\alpha\beta}\delta(\mathbf{r-r'})\delta(t-t')
\end{array}
\right.
\end{equation}
The dynamics is governed by a diffusion coefficient $D=1$ and a mobility $\sigma(\rho) = \rho^2$, related by the Einstein relation \cite{Bertini05},
\begin{equation}
    D = \sigma(\rho)\chi(\rho)^{-1} \, ,
\end{equation}
where $\chi(\rho)$ is the static compressibility. 

For time dependent processes, $\chi(\rho)$ is defined by (see section II.2.5 in \cite{Spohn})
\begin{equation}
    \chi(\rho(\mathbf{r})) = \int\limits_V d\mathbf{r'}S(\mathbf{r-r'},0) = \hat{S}(0,0) \, 
\label{chi}
\end{equation}
where $S(\mathbf{r-r'}, t-t')$ is the equilibrium correlation function of the density, $S(\mathbf{r-r'}, t-t') = \langle\rho(\mathbf{r},t)\rho(\mathbf{r'},t')\rangle - \langle\rho^2\rangle $. The Fourier transform $\hat{S}(\mathbf{k},t) = \int e^{i\mathbf{k}\cdot\mathbf{r}}S(\mathbf{r}, t)$ is known as the structure function or intermediate scattering function. 

For coherent light, the effective Langevin equation (7) is time independent. More precisely, it is equivalent to the KMP model integrated over the elastic mean time $\tau = \frac{l}{v}$. At time scales shorter than $\tau$, light behaves ballistically and this regime is integrated out in Eq.(7). The correspondence with the KMP model is obtained re-introducing a time dependence and by identifying $\rho(\mathbf{r},t)\equiv I(\mathbf{r},t)$ and $\mathbf{J}(\mathbf{r},t)\equiv \mathbf{j}(\mathbf{r},t)$ where $I(\mathbf{r},t)$ and $\mathbf{j}(\mathbf{r},t)$ are related by a time dependent Langevin equation
\begin{equation}
\mathbf{j}(\mathbf{r},t) = -D\bm{\nabla}I(\mathbf{r},t) + \tilde{\bm{\nu}}(\mathbf{r},t) \, ,
\end{equation}
where the noise term encapsulates the fluctuations created by both the coherent effects discussed in the main text and the ballistic trajectories. It is related to the noise term in Eq.(7) by
\begin{equation}
\langle\nu_{\alpha}(\mathbf{r})\nu_{\beta}(\mathbf{r'})\rangle = \int\limits_0^{\tau}dt\,  \langle\tilde{\nu}_{\alpha}(\mathbf{r},t)\tilde{\nu}_{\beta}(\mathbf{r'},0)\rangle \, .
\label{nu-time}
\end{equation}

\medskip

To derive the compressibility Eq.(10), we first note that Eq.(\ref{nu-time}) leads to
\begin{equation}
\langle\tilde{\nu}_{\alpha}(\mathbf{r},t)\tilde{\nu}_{\beta}(\mathbf{r'},t')\rangle =  \sigma\delta_{\alpha\beta}\delta(\mathbf{r-r'})\delta(t-t') \, .
\end{equation}
The time and space fluctuations are decoupled, hence
%
%
%
\begin{equation}
\begin{array}{ll}
\langle\delta I(\mathbf{r},t)\delta I(\mathbf{r'},t')\rangle & = \langle\delta I(\mathbf{r})\delta I(\mathbf{r'})\rangle\delta(t-t') \, ,
\end{array}
\end{equation}
and
\begin{equation}
\begin{array}{ll}
\langle\delta I(\mathbf{r})\delta I(\mathbf{r'})\rangle & = \int\limits_0^{\tau}dt\langle\delta I(\mathbf{r},t)\delta I(\mathbf{r'},t')\rangle \, .
\end{array}
\label{corr-i-t}
\end{equation}
From Eqs.(\ref{chi}, \ref{corr-i-t}), we obtain
\begin{equation}
\begin{array}{ll}
\int\limits_V d\mathbf{r'} \langle\delta I(\mathbf{r})\delta I(\mathbf{r'})\rangle &=\int\limits_V d\mathbf{r'} \int\limits_0^{\tau}dt\langle\delta I(\mathbf{r},t)\delta I(\mathbf{r'},t')\rangle\\
\\
& = \int\limits_0^{\tau}dt\chi(\mathbf{r})\\
\\
& = \tau \chi(\mathbf{r}) \, ,
\end{array}
\end{equation}
and we deduce the compressibility,
\begin{equation}
\begin{array}{ll}
     \chi(\mathbf{r}) &= \frac{1}{\tau}\int\limits_V d\mathbf{r'}\langle\delta I(\mathbf{r})\delta I(\mathbf{r'})\rangle^{(1)}\\
     \\
     &= \frac{1}{\tau}\frac{2\pi l}{k^2}I_D(\mathbf{r})^2\\
     \\
     &= \frac{2\pi v}{k^2}I_D(\mathbf{r})^2\\
     \\
     &= \frac{c_0}{D} I_D(\mathbf{r})^2 \, .
\end{array}
\end{equation}
Using Eq.(9), we recover the Einstein relation,
\begin{equation}
    \begin{array}{ll}
        \sigma(\mathbf{r})\chi(\mathbf{r})^{-1} = D \, .
    \end{array}
\end{equation}

\medskip

The Langevin approach is valid for weak noise, i.e. $\lim\limits_{V\to \infty}\sigma= 0$ over diffusive time scales $t\propto L^2$. To check that this requirement is met, it is useful to rescale the variables
\begin{equation}
\left\{
\begin{array}{ll}
\mathbf{x} = \frac{\mathbf{r}}{L}\\
\\
\tau = \frac{t}{L^2}
\end{array}
\right.
\label{units}
\end{equation}  
In these rescaled units,
\begin{equation}
\left\{
\begin{array}{ll}
I(\mathbf{x},\tau) = I(\mathbf{r},t)\\
\\
\mathbf{j}(\mathbf{x},\tau) = L\mathbf{j}(\mathbf{r},t)
\end{array}
\right.
\end{equation}
For the noise, we note that
\begin{equation}
\begin{array}{ll}
\langle\bm{\eta(r)\eta(r')}\rangle = \delta(\mathbf{r-r'}) \\
\\
= \delta(L(\mathbf{x-x'}))\\
\\
= \frac{1}{L^3}\delta(\mathbf{x-x'}) \\
\\
=\frac{1}{L^3}\langle\tilde{\bm{\eta}}(\mathbf{x})\tilde{\bm{\eta}}(\mathbf{x'})\rangle \, ,
\end{array}
\end{equation}
and $\bm{\eta}$ is therefore rescaled as $\bm{\eta}(\mathbf{x}) = \sqrt{L^3}\bm{\eta(r)}$.
In these units, the Langevin equation (7) becomes
\begin{equation}
\begin{array}{ll}
\frac{1}{L}\mathbf{j(x)} = -D\frac{1}{L}\nabla I(\mathbf{x}) + \sqrt{\frac{\sigma}{L^3}}\bm{\eta(x)}\\
\\
\mathbf{j(x)} = -D\nabla I(\mathbf{x}) + \sqrt{\frac{\sigma}{L}}\bm{\eta(x)}
\end{array}
\label{lan-light-res}
\end{equation}
and the noise term satisfies $\lim\limits_{L\to\infty}\frac{\sigma}{L}=0$. The Langevin approach is therefore valid.

\section{Fluctuation induced forces - scaling and exact expression in simple cases}
In this section, we first prove Eq.(12), namely that the light FIF in a scattering medium of finite volume $V$ illuminated by a monochromatic light source can be expressed in the universal form $ \langle \bm{f}^2\rangle = \frac{1}{g_{\mathcal{L}}}\frac{\mathcal{P}^2}{v^2}(\mathcal{Q}_2+\mathcal{Q_{\nu}})$. For simplicity we consider a cubic volume $V=L^3$, in which case $g_{\mathcal{L}} = \frac{k^2 l }{3\pi}L$, but the reasoning can be adapted for any geometry, e.g. in the geometry of the Fig.1 or Fig.\ref{setup-sm} where the volume $V$ is the rectangle parallelepiped contained between the plates. We then calculate $\mathcal{Q}_2, \mathcal{Q_{\nu}}$ exactly in two simple cases discussed in the main text, and show how to isolate $\bm{f_2}$ or $\bm{f_{\nu}}$ using appropriate boundary conditions.
\subsection{Universal scaling of the fluctuation induced forces}

In order to prove Eq.(12) we consider the two main contributions to the fluctuation induced forces, $\bm{f_2}$ and $\bm{f_{\nu}}$. The terms $\bm{f_1}$ and $\bm{f_3}$ are indeed negligible, since $\bm{f_1}^2\sim \left(\frac{l}{L}\right)^2\bm{f_{\nu}}^2$ and $\bm{f_3}^2\propto \frac{1}{g_{\mathcal{L}}^2}$. 

\medskip  

$\bullet$ \underline{Contribution of the intensity fluctuations}

\medskip

We derive here the term $\bm{f_2}$. By definition, 
\begin{equation}
\bm{f_2}^2 = \frac{D^2}{v^4} \oint\limits_{\partial V} d\mathbf{r}\oint\limits_{\partial V}d\mathbf{r'} (\mathbf{\hat{n}(r)}\cdot\boldsymbol{\nabla})(\mathbf{\hat{n}(r')}\cdot\boldsymbol{\nabla})\langle\delta I(\mathbf{r})\delta I(\mathbf{r'})\rangle^{(2)} \, .
\end{equation}
Using Eq.(\ref{corrI2}), we obtain:
\begin{align}
\bm{f_2}^2 &= c_0\frac{D^2}{v^4}\oint\limits_{\partial V} 
d\mathbf{r}\oint\limits_{\partial V} d\mathbf{r'}(\mathbf{\hat{n}(r)}\cdot\boldsymbol{\nabla})(\mathbf{\hat{n}(r')}\cdot\boldsymbol{\nabla}) \nonumber \\
& \int\limits_V d\mathbf{r_1}I_D(\mathbf{r_1})^2\boldsymbol{\nabla_{1}}P_D(\mathbf{r_1,r})\cdot\boldsymbol{\nabla_{1}}P_D(\mathbf{r_1,r'}) \, ,
\label{ti}
\end{align}
where $ \mathbf{\hat{n}(r)}$ is the local normal unitary vector at point $\mathbf{r}$ of the boundary $\partial V$. Note that the volume integral in Eq.(\ref{ti}) allows to understand qualitatively the limits $\lim\limits_{L\to 0}\bm{f_2}^2 = 0$ and $\lim\limits_{L\to \infty}\bm{f_2}^2 = 0$. The first limit is clear since the volume $V$ vanishes as $L\to 0$. The second limit results from the scaling of the functions $I_D(\mathbf{r})$ and $P_D(\mathbf{r,r'})$.
To study the scaling of $\bm{f_2}^2$, we change the integration variable to a dimensionless variable $\mathbf{u} = \mathbf{r}/L$, and we define
\begin{equation}
\left\{
\begin{array}{ll}
\tilde{P}_D(\mathbf{u,u'}) = DLP_D(\mathbf{r,r'}) \\
\\
\tilde{\gamma}(\mathbf{u}) = \frac{L^3}{v\mathcal{P}}\gamma'(\mathbf{r})
\end{array}
\right.
\label{scaled-fun}
\end{equation}
The new functions $\tilde{P}_D(\mathbf{u,u'})$ and $\tilde{\gamma}(\mathbf{u})$ are dimensionless, independent of the size $L$ of the system, and satisfy:
\begin{equation}
\left\{
\begin{array}{ll}
-\Delta_{\mathbf{u}} \tilde{P}_D(\mathbf{u,u'}) = \delta(\mathbf{u-u'})\\
\\
\int\limits_{\tilde{V}}d\mathbf{u}\tilde{\gamma}(\mathbf{u}) = 1
\end{array}
\right.
\end{equation}
where $\tilde{V} = [-1,1]^3$ is the rescaled volume.
With the rescaled functions in Eqs.(\ref{scaled-fun}), the surface integral in Eq.(\ref{ti}) becomes
\begin{align}
\oint\limits_{\partial V} 
d\mathbf{r}\oint\limits_{\partial V} d\mathbf{r'}(\mathbf{\hat{n}(r)}\cdot\boldsymbol{\nabla})(\mathbf{\hat{n}(r')}\cdot\boldsymbol{\nabla})\boldsymbol{\nabla_{1}}P_D(\mathbf{r_1,r})\cdot\boldsymbol{\nabla_{1}}P_D(\mathbf{r_1,r'}) \nonumber\\
= \frac{1}{D^2L^2}h(\mathbf{u_1})
\label{ti-int-1}
\end{align}
where $\mathbf{u_1} =\mathbf{r_1}/L $, and 
$$h(\mathbf{u_1}) = \oint\limits_{\partial \tilde{V}} 
d\mathbf{r}\oint\limits_{\partial \tilde{V}} d\mathbf{u'}(\mathbf{\hat{n}(u)}\cdot\boldsymbol{\nabla})(\mathbf{\hat{n}(u')}\cdot\boldsymbol{\nabla})\boldsymbol{\nabla_{1}}P_D(\mathbf{u_1},\mathbf{u})\cdot\boldsymbol{\nabla_{1}}P_D(\mathbf{u_1},\mathbf{u'})$$ 
is a dimensionless function of $\mathbf{r_1}/L$. Using Eqs.(\ref{scaled-fun}), we write $I_D^2(\mathbf{r_1})$ in the form
\begin{equation}
\begin{array}{ll}
I_D^2(\mathbf{r_1}) &= \iint\limits_{V\times V}d\mathbf{r}d\mathbf{r'}\gamma'(\mathbf{r})\gamma'(\mathbf{r'})P_D(\mathbf{r,r_1})P_D(\mathbf{r',r_1})\\
\\
&= \frac{v^2\mathcal{P}^2}{D^2L^2}\iint\limits_{\tilde{V}\times \tilde{V}}d\mathbf{u}d\mathbf{u'}\tilde{\gamma}(\mathbf{u})\tilde{\gamma}(\mathbf{u'})\tilde{P}_D(\mathbf{u,u_1})\tilde{P}_D(\mathbf{u',u_1})\\
\\
&= \frac{v^2\mathcal{P}^2}{D^2L^2}\tilde{I}_D^2(\mathbf{u_1}) \, ,
\end{array}
\label{ti-int-2}
\end{equation}
where $\tilde{I}_D(\mathbf{u_1}) = \int\limits_{\tilde{V}}d\mathbf{u}\tilde{\gamma}(\mathbf{u}))\tilde{P}_D(\mathbf{u,u_1})$ is a dimensionless function.
Replacing Eq.(\ref{ti-int-1}) and Eq.(\ref{ti-int-2}) in Eq.(\ref{ti}), and performing one last change of variable $\mathbf{u_1} = \mathbf{r_1}/L$, we obtain finally
\begin{equation}
\begin{array}{ll}
\bm{f_2}^2 &= c_0\frac{D^2}{v^4}\frac{1}{D^2L^2}\frac{v^2\mathcal{P}^2}{D^2L^2}\int\limits_V d\mathbf{r_1}\tilde{I}_D^2(\mathbf{r_1}/L)h(\mathbf{r_1}/L)\\
\\
&= \frac{\mathcal{P}^2}{v^2}\frac{c_0}{D^2L^4}L^3\int\limits_{\tilde{V}} d\mathbf{u_1}\tilde{I}_D^2(\mathbf{u_1})h(\mathbf{u_1})\\
\\
&=  \frac{\mathcal{P}^2}{v^2}\frac{3\pi}{k^2 l L}2\int\limits_{\tilde{V}} d\mathbf{u_1}\tilde{I}_D^2(\mathbf{u_1})h(\mathbf{u_1})\\
\\
&= \frac{1}{g_{\mathcal{L}}}\frac{\mathcal{P}^2}{v^2}\mathcal{Q}_2 \, ,
\end{array}
\end{equation}
where 
$$\mathcal{Q}_2 = 2\int\limits_{\tilde{V}} d\mathbf{u_1}\tilde{I}_D^2(\mathbf{u_1})h(\mathbf{u_1})$$
is a dimensionless number, which depends on the shape of the system and of the light source, but which is independent of the size $L$ of the system.\\

\medskip  

$\bullet$ \underline{Contribution of the noise}

\medskip

We derive here the noise contribution $\bm{f_{\nu}}^2$. By definition,
\begin{equation}
\bm{f_{\nu}}^2 = \frac{1}{v^4}\oint\limits_{\partial V}d\mathbf{r}\oint\limits_{\partial V}d\mathbf{r'}\langle\bm{\nu(r)\nu(r')}\rangle \, .
\end{equation}
Using Eq.(8),
\begin{equation}
\bm{f_{\nu}}^2=\frac{1}{v^4}\frac{c_0}{L}\oint\limits_{\partial V}d\mathbf{r}I_D(\mathbf{r})^2 \, .
\label{tc}
\end{equation}
We perform the same change of variables and functions as in the previous paragraph, in Eqs.(\ref{scaled-fun}). Using Eq.(\ref{ti-int-2}), we obtain
\begin{equation}
\begin{array}{ll}
\bm{f_{\nu}}^2 &= \frac{c_0}{v^4L}\frac{v^2\mathcal{P}^2}{D^2L^2}L^2\oint\limits_{\partial \tilde{V}}d\mathbf{u}\tilde{I}_D(\mathbf{u})^2\\
\\
& = \frac{1}{g_{\mathcal{L}}}\frac{\mathcal{P}^2}{v^2}\mathcal{Q}_{\nu}
\end{array}
\end{equation}
with 
$$\mathcal{Q}_{\nu} = 2\oint\limits_{\partial \tilde{V}}d\mathbf{u}\tilde{I}_D(\mathbf{u})^2$$
a dimensionless number, which depends on the shape of the system and of the light source, but not on the size $L$ of the system.

\medskip
In conclusion, we proved that the fluctuation induced forces are of the form given in Eq.(12),
\begin{equation}
\langle \bm{f}^2\rangle = \bm{f_{\nu}}^2 + \bm{f_2}^2 = \frac{1}{g_{\mathcal{L}}}\frac{\mathcal{P}^2}{v^2}\left(\mathcal{Q}_2+\mathcal{Q}_{\nu}\right) \, ,
\end{equation}
with $\mathcal{Q}_2$, $\mathcal{Q}_{\nu}$ dimensionless numbers, depending on the shape of the system, on the boundary conditions and on the light source, but not of the size $L$. 
%
\subsection{Exact expression of the fluctuation induced forces in a simple case}
\begin{figure}
\centering
\hspace*{-2cm}\includegraphics[scale=0.6]{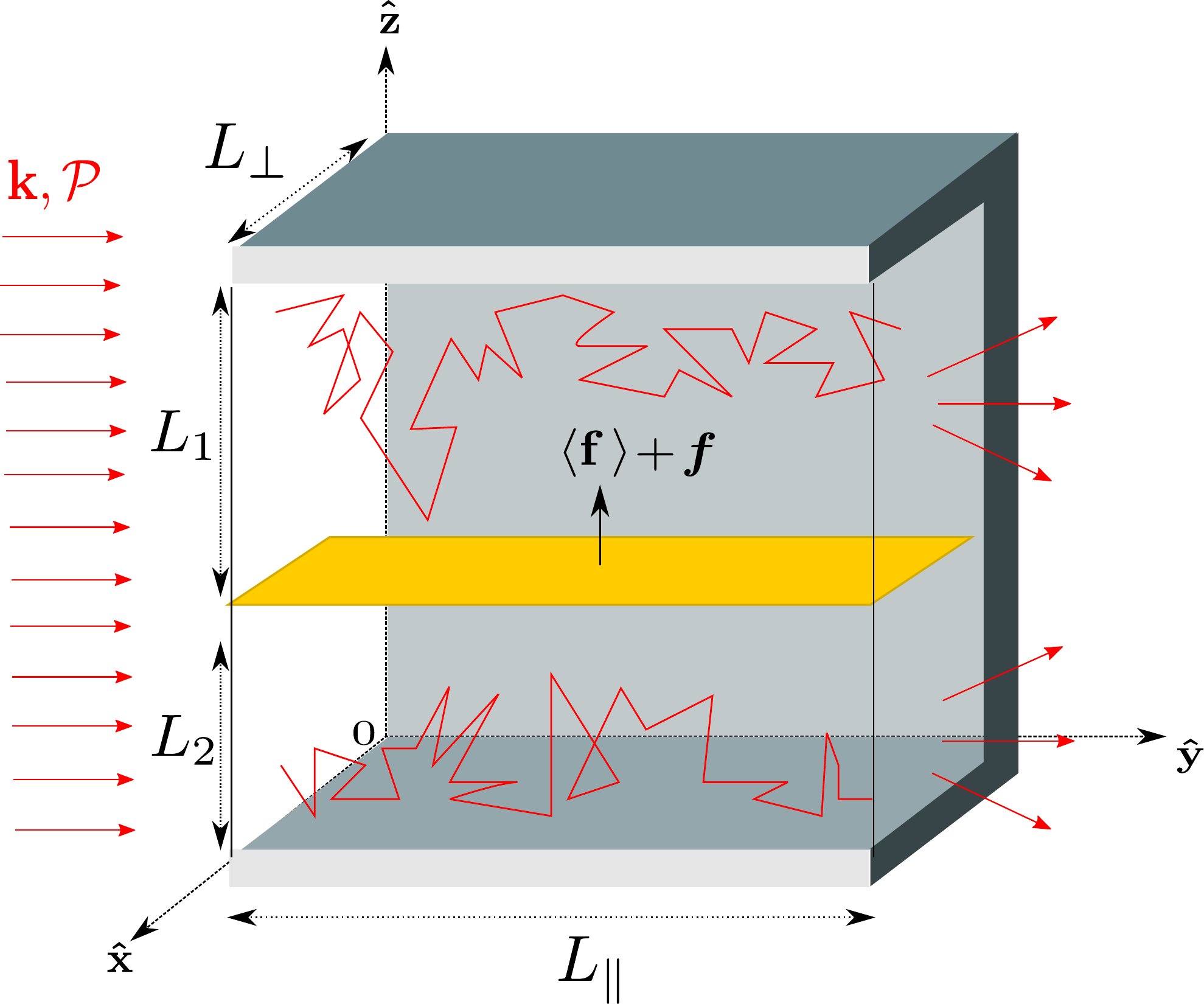}
\caption{A scattering medium contained in a box of size $L_{\parallel}\times L_{\perp}\times (L_1+L_2)$ is illuminated by a monochromatic plane wave. A plate (yellow) separates the medium in two zones, labeled by $j=1,2$, of thicknesses $L_j$. The two zones have a priori different conductances. In the particular case $L_1=L_2$, the average radiative force on the plate vanishes to zero, which is not true a priori in the general case - except for perfectly reflecting plates (see text). }
\label{setup-sm}
\end{figure}

We derive here the exact analytical expression of the fluctuation induced forces for the simple setup Fig.\ref{setup-sm}, which is a generalization of the case illustrated in Fig.1 (in the latter, $L_1=L_2$). We also show that one can isolate the terms $\bm{f_{\nu}}$ or $\bm{f_2}$ using different boundary conditions: with a reflecting plate,  $\langle\bm{f}^2\rangle = \bm{f_{\nu}}^2$, while with an absorbing plate, $\langle\bm{f}^2\rangle = \bm{f_2}^2$.

A scattering medium is contained in a box $L_{\parallel}\times L_{\perp}\times (L_1+L_2)$. It is illuminated with a monochromatic plane wave propagating in the $\mathbf{\hat{y}}$ direction and emitted by a light source located outside of the medium, which means that $\gamma=0$ in the diffusion equation Eq.(\ref{diff-eq}). A thin plate or membrane of surface $S = L_{\perp}\times L_{\parallel}$ (yellow in Fig.\ref{setup-sm}) is placed inside the box, which splits the box into two zones, denoted $j=1,2$, respectively above and under the plate (see Fig.\ref{setup-sm}). As in the main text, we impose Neumann boundary conditions on the slabs at $y=0,L_{\parallel}$ and Dirichlet boundary conditions on the slabs at $x=0,L_{\perp}$. Those boundary conditions apply to the specific intensity. They translate, for $I_D(\mathbf{r})$ and $P_D(\mathbf{r,r'})$, to Neumann boundary conditions $y=0,L_{\parallel}$ and Dirichlet boundary conditions at $x=-l_0,L_{\perp}+l_0$ with $l_0=\frac{2l}{3}$ (see \cite{Akkermans}),
\begin{equation}
\left\{
\begin{array}{ll}
\mbox{Dirichlet: } P_D(\mathbf{r,r'})=0 \mbox{ for all $\mathbf{r}=(x,y,z)$ s.t. $x=-l_0,L_{\perp}+l_0$}\\
\\
\mbox{Neumann: }\partial_{y}P_D(\mathbf{r,r'})=0 \mbox{ for all $\mathbf{r}=(x,y,z)$ s.t. $y=0,L$}
\end{array}
\right.
\label{bc}
\end{equation}
%
We note $S_{\perp} = L_{\perp}(L_1+L_2)$ the illuminated surface, and we introduce the intensity of the light source, assumed to be uniform,
\begin{equation}
I = \frac{\mathcal{P}}{L_{\perp}(L_1+L_2)} = \frac{\mathcal{P}}{S_{\perp}} \, .
\end{equation}
We consider the cases of a reflective plate and of an absorbing one. In each case, we write the FIF in the general form Eq.(12), and we give the exact expression of $g_{\mathcal{L}}$. In the geometry under study here, where the volumes delimited by the plates are rectangular boxes, $g_{\mathcal{L}}$ is defined by adapting the Thouless argument or equivalently by using the reasoning in the chapter 12 in \cite{Akkermans}, namely, in each zone $j=1,2$,
\begin{equation}
g_{\mathcal{L}}^{(j)} \equiv \frac{k^2 l V_j}{3\pi (\max(L_j^2,L_{\perp}^2,L_{\parallel}^2))} \, ,
\label{gc}
\end{equation}
where $V_j = L_jL_{\parallel}L_{\perp}$.
%

\medskip

$\bullet$\underline{Reflecting plate}

\medskip 

\begin{figure}
\centering
\vspace*{-7cm}
\hspace*{-2cm}\includegraphics[scale=0.7]{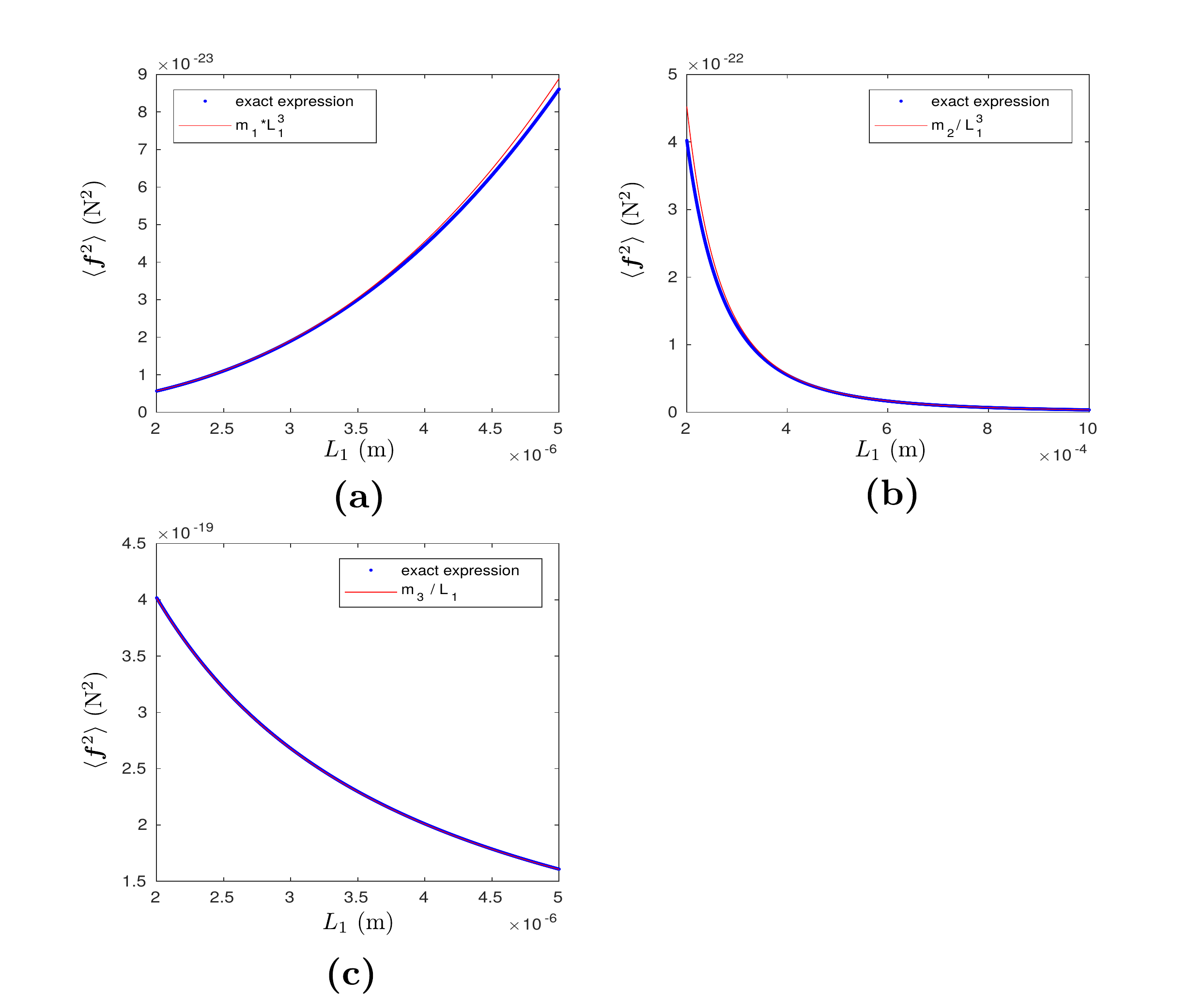}
\caption{Asymptotic behavior and fitted functions of $\langle \bm{f}^2\rangle$ with Neumann boundary conditions in the direction of the illuminating light beam and Dirichlet boundary conditions in the direction $\mathbf{\hat{x}}$, for an absorbing plate {\bf (a)}, {\bf (b)} and a reflecting plate {\bf (c)}. We assume $L_2\gg L_1$ for simplicity; we therefore neglect the FIF from the zone 2 in Fig.\ref{setup-sm}. We fix $L_{\parallel}=L_{\perp} = 40\, \mu $m, as in Table I in the main text. {\bf (a)} For $L_1\ll L_{\parallel},L_{\perp}$, we have $ g_{\mathcal{L}}^{(1)}=\frac{k^2l L_1L_{\perp}}{3\pi L_{\parallel}}$, and $\langle \bm{f}^2\rangle$ scales like $L_1^3$. The fit parameter $m_1$ numerically matches the value obtained by taking $L_1\ll L_{\parallel},L_{\perp}$ in Eq.(\ref{dfperp-pw-d-plate-n-in}).{\bf (b)} For $L_1\gg L_{\parallel},L_{\perp}$, $ g_{\mathcal{L}}^{(1)}=\frac{k^2 l L_{\parallel}L_{\perp}}{3\pi L_1}$ and the fluctuating force scales like $1/L_1^3$, with a fit factor $m_2$ also matching the theoretical value expected in the limit $L_1\gg L_{\parallel},L_{\perp}$. {\bf (c)} In the case of reflecting plates, the force scales like $1/L_1$ at all length scales, with $m_3$ matching the value expected from Eq.(\ref{dfperpn}).}
\label{fit}
\end{figure}

With a reflecting plate, $\partial_z P_D(\mathbf{r,r'}) = 0$ for all $\mathbf{r}$ on the plates. From Eq.(\ref{ti}), we deduce that $\bm{f_2}^2 = 0$, i.e. that the intensity fluctuations will not contribute to the fluctuation induced forces. \\
The Green's function $P_D(\mathbf{r,r'})$ is

\begin{equation}
\begin{array}{ll}
P_D(\mathbf{r,r'}) = \sum\limits_{n_1>0,n_2,n_3\geq 0}\frac{\psi_{n_1 n_2 n_3}(\mathbf{r})\psi_{n_1 n_2 n_3}(\mathbf{r'})}{E_{n_1 n_2 n_3}} \, ,
\end{array}
\label{pd-n}
\end{equation}
where, for $j=1,2$, 
\begin{equation}
\psi_{n_1 n_2 n_3}(\mathbf{r}) = N\sin\left(\frac{n_1\pi (x+l_0)}{L_{\perp}+2l_0}\right)\cos\left(\frac{n_2\pi y}{L_{\parallel}}\right)\cos\left(\frac{n_3\pi z}{L_{j}}\right)
\label{eig-n}
\end{equation}
are the eigenfunctions of the diffusion equation with the boundary conditions considered, normalized with $N=\sqrt{\frac{2^3}{L_{\parallel}L_jL_{\perp}}}$, and with eigenvalues 
\begin{equation}
E_{n_1 n_2 n_3} = D\pi^2\left(\frac{n_1^2}{L_{\perp}^2}+\frac{n_2^2}{L_{\parallel}^2}+\frac{n_3^2}{L_{j}^2}\right) \, .
\end{equation}
We deduce $I_D(\mathbf{r})$ using Eq.(\ref{diff-eq}) and the Green's identity Eq.(\ref{green}). Since there is no light source inside the medium, the right hand side of the diffusion equation Eq.(\ref{diff-eq}) is reduced to $vI_0(\mathbf{r})/l$. We solve Eq.(\ref{eq-diff-i0}) to obtain $I_0(\mathbf{r})$,
\begin{equation}
I_0(\mathbf{r})=Ie^{-y/l} \, .
\end{equation}
We then obtain for $I_D(\mathbf{r})$,
\begin{equation}
I_D(\mathbf{r}) = \frac{3DI}{l^2}\int\limits_V e^{-y'/l}P_D(\mathbf{r,r'})d\mathbf{r'} \, .
\label{id-n}
\end{equation}
The average intensity $I_D(\mathbf{r})$ is independent of $z$, which means that the average force on the plate is equal to zero. \\
As we mentioned at the beginning of the paragraph, with reflecting plates $\bm{f_2}^2 = 0$ and the main contribution comes from the noise term $\bm{f_{\nu}}^2$ alone. Re-injecting Eq.(\ref{pd-n}) and Eq.(\ref{id-n}) in Eq.(\ref{tc}), we obtain, after a standard calculation, the fluctuation induced force on the plate,
 \begin{equation}
 \begin{array}{ll}
\langle \bm{f }^2\rangle & = b\frac{I^2}{v^2}\frac{3\pi L_{\parallel}}{k^2 l L_{\perp}}\left(\frac{1}{L_1}+\frac{1}{L_{2}}\right)S_{\parallel}^2\beta'\left(\frac{L_{\parallel}}{L_{\perp}}\right)\\
\\
& = b\frac{\mathcal{P}^2}{v^2}\frac{3\pi L_{\parallel}}{k^2 l L_{\perp}}\left(\frac{1}{L_1}+\frac{1}{L_{2}}\right)\beta\left(\frac{L_{\parallel}}{L_{\perp}},\frac{S_{\parallel}}{S_{\perp}}\right) \, ,
 \end{array}
\label{dfperpn}
\end{equation}
where $b\sim 0.13$ is a prefactor englobing a numerical fit factor and the normalization factor of the eigenfunctions in Eq.(\ref{eig-n}). The term $\beta\left(\frac{L_{\parallel}}{L_{\perp}},\frac{S_{\parallel}}{S_{\perp}}\right)$ is a dimensionless function of the aspect ratios $\frac{L_{\parallel}}{L_{\perp}}$, $\frac{S_{\parallel}}{S_{\perp}}$, equal to
\begin{equation}
 \beta\left(\frac{L_{\parallel}}{L_{\perp}},\frac{S_{\parallel}}{S_{\perp}}\right) = \frac{S_{\parallel}^2}{S_{\perp}^2}\sum_{n_2\geq 0,n_1\mbox{\tiny{odd}}}\frac{1}{n_1^2\left( n_2^2+n_1^2\frac{L_{\parallel}^2}{L_{\perp}^2}  \right)^2} \, .
\end{equation}
The expression in Eq.(\ref{dfperpn}) for the fluctuation induced forces holds for any value of the parameters $L_{\parallel},L_{\perp},Lj$. A plot of this FIF is given on Fig.\ref{fit}.c. To write Eq.(\ref{dfperpn}) in the form of the Eq.(12), we need to define $g_{\mathcal{L}}^{(j)}$ explicitly and check that $g_{\mathcal{L}}^{(j)}\gg 1$. 
According to Eq.(\ref{gc}), if $L_{\parallel}\simeq L_{\perp}$ and $L_{\parallel}, L_{\perp} \gg L_j$ then $g_{\mathcal{L}}^{(j)} = \frac{k^2l L_jL_{\perp}}{3\pi L_{\parallel}}$ which leads to
\begin{equation}
 \langle \bm{f }^2\rangle = \frac{\mathcal{P}^2}{v^2}\left(\frac{1}{ g_{\mathcal{L}}^{(1)}}+\frac{1}{ g_{\mathcal{L}}^{(2)}}\right) \mathcal{Q}_{\nu}\left(\frac{L_{\parallel}}{L_{\perp}},\frac{S_{\parallel}}{S_{\perp}}\right) \, ,
\end{equation}
with $\mathcal{Q}_{\nu}\left(\frac{L_{\parallel}}{L_{\perp}},\frac{S_{\parallel}}{S_{\perp}}\right) = b\beta\left(\frac{L_{\parallel}}{L_{\perp}},\frac{S_{\parallel}}{S_{\perp}}\right)$.\\
This expression is valid provided that $g_{\mathcal{L}}^{(j)}\gg 1$, i.e. $\frac{k^2l L_jL_{\perp}}{3\pi L_{\parallel}}\gg 1$. In the main text and in Fig.\ref{fit}, we choose $l = 1\, \mu$m and $kl = 10$. This choice leads to the requirement: $L_j\gg 10^{-7}$m, which is readily satisfied both in the letter and Fig.\ref{fit}.\\
On the other hand, for $L_j\gg L_{\parallel},L_{\perp}$, then $g_{\mathcal{L}}^{(j)} = \frac{k^2l L_{\parallel}L_{\perp}}{3\pi L_j}$ and the FIF can be written in the form
\begin{equation}
 \langle \bm{f }^2\rangle = \frac{\mathcal{P}^2}{v^2}\left(\frac{1}{ g_{\mathcal{L}}^{(1)}}\mathcal{Q}_{\nu}\left(\frac{L_{\parallel}}{L_1},\frac{L_{\parallel}}{L_{\perp}}\right)+\frac{1}{ g_{\mathcal{L}}^{(2)}}\mathcal{Q}_{\nu}\left(\frac{L_{\parallel}}{L_2},\frac{L_{\parallel}}{L_{\perp}}\right)\right) \, ,
\end{equation}
with $\mathcal{Q}_{\nu}\left(\frac{L_{\parallel}}{L_j},\frac{L_{\parallel}}{L_{\perp}},\frac{S_{\parallel}}{S_{\perp}}\right) = b\frac{L_{\parallel}^2}{L_j^2}\beta\left(\frac{L_{\parallel}}{L_{\perp}},\frac{S_{\parallel}}{S_{\perp}}\right)$.
In that case, the requirement $g_{\mathcal{L}}^{(j)}\gg 1$ translates to $L_j\ll 10^{7}L_{\parallel}L_{\perp}$m. In the letter and Fig.\ref{fit}, we choose $L_{\parallel}=L_{\perp} =40\, \mu $m, therefore $g_{\mathcal{L}}^{(j)}\gg 1$ means that we should have $L_j\ll 10^{-2}$m, which is the case in all the numerical results presented. 

\medskip

$\bullet$\underline{Absorbing plate}

\medskip 

With an absorbing plate, $I_D(\mathbf{r}) = 0$ for all $\mathbf{r}$ on the plates \footnote{The Dirichlet boundary condition expressed in Eq.(\ref{bc}) is valid at an interface between a scattering and non scattering medium; since the plate is immersed in the medium, the Dirichlet boundary condition is simply formulated as $I_D(\mathbf{r}) = 0$ }. From Eq.(\ref{tc}), we obtain $\bm{f_{\nu}}^2 = 0$, i.e. contrary to the previous case of a reflecting plate, here the intensity fluctuations are the main contribution to the fluctuation induced forces. \\
The Green's function is
\begin{equation}
P_D(\mathbf{r,r'}) = \sum\limits_{n_1,n_3> 0, n_2 \geq 0}\frac{\psi_{n_1 n_2 n_3}(\mathbf{r})\psi_{n_1 n_2 n_3}(\mathbf{r'})}{E_{n_1 n_2 n_3}}
\end{equation} 
with
\begin{equation}
\psi_{n_1 n_2 n_3}(\mathbf{r}) = N\sin\left(\frac{n_1\pi (x+l_0)}{L_{\perp}+2l_0}\right)\cos\left(\frac{n_2\pi y}{L_{\parallel}}\right)\sin\left(\frac{n_3\pi z}{L_j}\right)
\label{eig-d}
\end{equation}
$N=\sqrt{\frac{2^3}{L_{\parallel}L_jL_{\perp}}}$, and with $E_{n_1 n_2 n_3} = D\pi^2\left(\frac{n_1^2}{L_{\perp}^2}+\frac{n_2^2}{L_{\parallel}^2}+\frac{n_3^2}{L_{j}^2}\right)$. \\
The normal force on the plate is equal to
\begin{equation}
\langle \mathbf{f}\rangle = 
\frac{I}{v}\frac{2^6}{\pi^4}L_{\perp}
\left[\sum\limits_{n_1,n_3 \mbox{ \tiny{odd}}}L_1\frac{1}{n_1^2\left(n_3^2+\frac{n_1^2 L^2_1}{L^2_{\perp}}\right)}-L_2\frac{1}{n_1^2\left(n_3^2+\frac{n_1^2 L^2_2}{L^2_{\perp}}\right)}\right]\mathbf{\hat{z}} 
\end{equation}
and is well approximated by
\begin{equation}
\langle \mathbf{f}\rangle\simeq \frac{I}{v}\frac{2^6}{\pi^4}L_{\perp}\left(\frac{1}{L_1}\frac{1}{\left(\frac{1}{L_1^2}+\frac{1}{L_{\perp}^2}\right)}-\frac{1}{L_2}\frac{1}{\left(\frac{1}{L_2^2}+\frac{1}{L_{\perp}^2}\right)}\right)\mathbf{\hat{z}} \, .
\label{f1-abs}
\end{equation}
It is straightforward from Eq.(\ref{f1-abs}) that the average force cancels out when $L_1 = L_2$.
Re-injecting the expressions of $I_D(\mathbf{r})$, $P_D(\mathbf{r,r'})$ in Eq.(\ref{ti}), we obtain the fluctuation induced force on $p_1$,
\begin{equation}
\begin{array}{ll}
\langle \bm{f }^2\rangle = a\frac{I^2}{v^2}\frac{3\pi}{k^2 l}\left[\frac{L_1 L_{\perp}^5}{L_{\parallel}^3}\delta\left(\frac{L_1}{L_{\parallel}},\frac{L_{\perp}}{L_{\parallel}}\right)+\frac{L_2 L_{\perp}^5}{L^3}\delta\left(\frac{L_2}{L_{\parallel}},\frac{L_{\perp}}{L_{\parallel}}\right)\right] \, ,
\end{array}
\label{dfperp-pw-d-plate-n-in}
\end{equation}
with $a\sim 0.3$ a prefactor englobing a fit factor and the normalization of the eigenfunctions Eq.(\ref{eig-d}), and $\delta\left(\frac{L_j}{L_{\parallel}},\frac{L_{\perp}}{L_{\parallel}}\right)$ a dimensionless geometrical correction. The exact expression of $\delta\left(\frac{L_j}{L_{\parallel}},\frac{L_{\perp}}{L_{\parallel}}\right)$ is quite heavy but can be obtained by a standard calculation using Eq.(\ref{ti}). To alleviate the discussion, we do not give its full expression here. After performing the integrals in Eq.(\ref{ti}), we can write $\delta\left(\frac{L_j}{L_{\parallel}},\frac{L_{\perp}}{L_{\parallel}}\right)$ as a product of two sums, which can be well approximated by keeping the first terms,
\begin{equation}
\begin{array}{ll}
\delta\left(\frac{L_j}{L_{\parallel}},\frac{L_{\perp}}{L_{\parallel}}\right) \sim
 \frac{1}{\left(\frac{L_j^2}{L_{\parallel}^2}+\frac{L_{\perp}^2}{L_{\parallel}^2}\right)^2}\left[\frac{1}{\left(1+\frac{L_{\parallel}^2}{L_j^2}+\frac{L_{\parallel}^2}{L_{\perp}^2}\right)^2}+\frac{1}{\frac{L_{\parallel}^2}{L_j^2}+\frac{L_{\parallel}^2}{L_{\perp}^2}}\right] \, .
\end{array}
\end{equation}
Note that the light FIF in this case tends to zero in both limits $L_1\to 0$ and $L_1\to \infty$; a plot of these two limits is given on Fig.\ref{fit}.a and Fig.\ref{fit}.b. A maximum is reached when $L_1\simeq L_{\perp}$. Note also that the FIF increase as $L_{\parallel}$ decreases. In the configuration where $L_j= L_{\perp}\gg L_{\parallel}$, we have $g_{\mathcal{L}}^{(j)} = \frac{k^2 l L_{\parallel}L_{\perp} }{3\pi L_j}$ and the FIF can be written in the form
\begin{equation}
\langle \bm{f }^2\rangle = \frac{\mathcal{P}^2}{v^2}\left(\frac{1}{ g_{\mathcal{L}}^{(1)}}\mathcal{Q}_2\left(\frac{L_1}{L_{\parallel}},\frac{L_{\perp}}{L_{\parallel}},\frac{S_{\perp}}{S_{\parallel}}\right)+\frac{1}{ g_{\mathcal{L}}^{(2)}}\mathcal{Q}_2\left(\frac{L_2}{L_{\parallel}},\frac{L_{\perp}}{L_{\parallel}},\frac{S_{\perp}}{S_{\parallel}}\right)\right)
\end{equation}
with $\mathcal{Q}_2\left(\frac{L_j}{L_{\parallel}},\frac{L_{\perp}}{L_{\parallel}},\frac{S_{\perp}}{S_{\parallel}}\right) = a\frac{L_{\perp}^4}{L_{\parallel}^2}\frac{S_{\parallel}^2}{S_{\perp}^2}\delta\left(\frac{L_j}{L_{\parallel}},\frac{L_{\perp}}{L_{\parallel}}\right)$.


%
\newpage
\renewcommand\refname{Bibliography}
\bibliographystyle{unsrtnat} 
\bibliography{casimir.bib}{} 